\documentclass[11pt]{article}
\newcommand{\beq}{\begin{equation}}
\newcommand{\eeq}{\end{equation}}
\newcommand{\beqa}{\begin{eqnarray}}
\newcommand{\eeqa}{\end{eqnarray}}

\newcommand{\sn}{{\rm sn}}
\newcommand{\cn}{{\rm cn}}
\newcommand{\dn}{{\rm dn}}
\newcommand{\ds}{{\rm ds}}
\newcommand{\cs}{{\rm cs}}
\newcommand{\ns}{{\rm ns}}
\newcommand{\cd}{{\rm cd}}

\newcommand{\sech}{{\rm sech}}
\newcommand{\sss}{{\vspace{.2in}}}

\begin{document}
\vspace{.2in}
~\hfill{\footnotesize IOP-BBSR/04-11}
\vspace{.5in}
\begin{center}
{\LARGE {\bf Cyclic Identities Involving Ratios of Jacobi Theta Functions}}
\end{center}
\vspace{.5in}
\begin{center}
{\Large{\bf  \mbox{Avinash Khare}
 }}\\
\noindent
{\large Institute of Physics, Sachivalaya Marg, Bhubaneswar 751005, India}
\end{center}
\vspace{.2in}
\begin{center}
{\Large{\bf  \mbox{Arul Lakshminarayan}
 }}\\
\noindent
{\large Department of Physics, IIT Madras, Chennai 600036, India}
\end{center}
\vspace{.2in}
\begin{center}
{\Large{\bf  \mbox{Uday Sukhatme}
 }}\\
\noindent
{\large Department of Physics, State University of New York at
Buffalo, Buffalo, NY 14260, U.S.A.}\\
\end{center}
\vspace{1.2in}

{\bf {Abstract:}}
Identities involving cyclic sums of terms composed from Jacobi elliptic 
functions evaluated at $p$ equally shifted points were recently found. The purpose of this paper is to re-express these cyclic identities in terms of ratios of Jacobi theta functions,
since many physicists prefer using Jacobi theta functions rather than Jacobi elliptic functions. 

\newpage
\section{Introduction.}

Recently, we have discovered many new cyclic identities involving the Jacobi elliptic functions
$\sn \,(x,m)$, $\cn \,(x,m)$, $\dn \,(x,m)$, where $m$ is the elliptic
modulus parameter $( 0\leq m\leq 1)$. These mathematical identities are described in references \cite{ksjmp,klsjmp,klsmp}, henceforth referred to as I, II and IIa 
respectively. The
functions $\sn \,(x,m)$, $\cn \,(x,m)$, $\dn \,(x,m)$ are doubly periodic
functions with periods $(4K(m), i2K'(m))$, $(4K(m), 2K(m)+i2K'(m))$, 
$(2K(m), i4K'(m))$,
respectively \cite{abr}.
Here, $K(m)$ denotes the complete elliptic integral of the first kind, and
$K'(m)= K(1-m)$. The $m =0$ limit gives
$K(0)=\pi /2$ and trigonometric functions: $\sn(x,0)=\sin x, ~\cn(x,0)=\cos x,
~\dn(x,0)=1$. The $m \rightarrow 1$ limit gives $K(1) \rightarrow \infty$ and
hyperbolic functions: $\sn(x,1) \rightarrow \tanh x,
~\cn(x,1) \rightarrow \sech \,x, ~\dn(x,1) \rightarrow \sech\, x$. 
For simplicity, from now on
we will not explicitly display the modulus parameter $m$ as an argument of the
Jacobi elliptic functions.

The cyclic identities
discussed in I and II play an important role in showing
that a kind of linear superposition is valid for many nonlinear differential
equations of physical interest \cite{ksprl,cks}. In all identities, the
arguments of the Jacobi functions in successive terms
are separated by either $2K(m)/p$ or $4K(m)/p$, where $p$ is an integer.
Each $p$-point
identity of rank $r$ involves a cyclic homogeneous polynomial of degree
$r$ (in Jacobi elliptic functions with $p$ equally spaced arguments) related
to other cyclic homogeneous  polynomials of degree $r-2$ or smaller.

In II, it was shown that all our identities follow from four master identities.
It was also shown that corresponding to every such identity, one can
obtain new identities corresponding to pure imaginary shifts by
multiples of $i2K'(m)/p$ or $i4K'(m)/p$, as well as identities
corresponding to complex shifts by multiples of $2[K(m)+iK'(m)]/p$
or $4[K(m)+iK'(m)]/p$. Identities involving the nine
secondary Jacobi elliptic functions [$\cd \,(x,m)$, $\ns \,(x,m)$, $\ds \,(x,m)$, etc.] were also discussed.
Furthermore, in II we gave results for several identities involving Weierstrass 
elliptic functions and ratios of Jacobi
theta functions, both of which are
intimately related with Jacobi elliptic functions \cite{abr}.

In particular, in II we showed that given any identity for the Jacobi elliptic
functions, we can immediately write down the corresponding identity for the
ratio of Jacobi theta functions, since the ratio of any two Jacobi theta functions 
is also doubly periodic. 
In discussions with several physicists, it became apparent that they are more comfortable thinking in terms of Jacobi theta
functions rather than Jacobi elliptic functions. In order for them to fully appreciate the power of various cyclic identities previously obtained
in I, II and  IIa, we decided to re-express these identities  
in terms of the ratios of Jacobi theta functions. 

In this context it is worth noting that the connection between the four 
Jacobi theta functions 
$\theta_1(z, \tau), \theta_2(z, \tau), \theta_3(z, \tau), \theta_4(z, \tau)$ 
and the Jacobi elliptic 
functions is given by \cite{law}
\beq\label{1}
\sn (u, m)=\frac{1}{m^{1/4}}\frac{\theta_1(z, \tau)}{\theta_4(z, \tau)}~,~~
\cn (u, m)=\frac{(1-m)^{1/4}}{m^{1/4}}\frac{\theta_2(z, \tau)}
{\theta_4(z, \tau)}~,~~
\dn (u, m)={(1-m)^{1/4}}\frac{\theta_3(z, \tau)}{\theta_4(z, \tau)}~,
\eeq
where $z \equiv \frac{u\pi}{2K(m)}$ and $\tau = iK'(m)/K(m)$. 
Therefore, any of our cyclic identities for real, imaginary or complex shifts
can also be re-written as identities for the ratios of Jacobi theta 
functions for shifts in units of $\pi /p$, $\pi \tau /p$ or $\pi (1+\tau)/p$
respectively if 
the ratios of theta functions involved are of period $\pi$ (or twice these values if the period is $2\pi$).
In particular note that while $\theta_{3,4} (z)$ are of period $\pi$,
$\theta_{1,2} (z)$ are of period $2\pi$.

For simplicity, from now on 
we will not explicitly display $\tau$ as an argument of the
Jacobi theta functions. We also point out that the constants
$m,1-m$ and $K(m)$ can also be entirely re-expressed 
in terms of $theta$ functions:
\beq\label{2}
m^{1/4} = \frac{\theta_2 (0)}{\theta_3 (0)}~,~(1-m)^{1/4} 
= \frac{\theta_4 (0)}{\theta_3 (0)}~,~\frac{2K(m)}{\pi} = \theta_3 ^2 (0)~.
\eeq
Further, the Jacobi zeta function $Z(u)$ 
and the complete elliptic integral of the second kind $E$, 
which also appear in some of the identities,  
can also be expressed in terms of Jacobi theta functions:
\beq\label{2a}
Z(u) = \frac{1}{\theta_3^2 (0)}\frac{\theta_4'(z)}
{\theta_4 (z)}~,~E = [1-\frac{\theta_4''(0)}
{\theta_3^4 (0) \theta_4 (0)}]K=
[1-\frac{\theta_4''(0)}
{\theta_3^4 (0) \theta_4 (0)}] \pi \theta_3^2 (0)\,.
\eeq
Finally, let us note a remarkable fact about the cyclic identities:
\cite{klspr}
\beq\label{2aa}
\sum_{j=1}^{p} g(z_j)[h(z_{j+1})\pm h(z_{j-1})]=
\pm \sum_{j=1}^{p} h(z_j)[g(z_{j+1})\pm g(z_{j-1})]\,,
\eeq
\beq\label{2aaa}
\sum_{j=1}^{p} (-1)^{j} g(z_j)[h(z_{j+1})\pm h(z_{j-1})]=
\mp \sum_{j=1}^{p} (-1)^{j} h(z_j)[g(z_{j+1})\pm g(z_{j-1})]\,,
\eeq
where $h(z)$ and $g(z)$ are combinations of ratios of Jacobi 
theta functions. Hence we shall only mention one of the two 
cyclic identities in order to avoid duplication.

The plan of this paper is as follows. We first consider the cyclic 
identities in 
I and II following from master identities MI-I to MI-IV. In
Sec. 2 to Sec. 5, we give a list of corresponding cyclic identities in 
terms of the ratios of Jacobi theta functions.
Finally, in Sec. 6 we 
summarize the results obtained and 
also indicate how to generalize these results to shifts in units of 
$T \tau /p$ as well as $T (1+\tau) /p$ (where the period $T$ is $\pi$ or 
$2\pi$). Further, we  also indicate how to obtain identities for other ratios of $\theta$ functions as well as for their products.

Throughout this paper, we use the notation  
$ 1 \le r < p$ and $(r,p) =1$, i.e. they are 
co-prime. 
It may be noted that for MI-I and MI-II identities
the period for the ratios of $theta$ functions is $\pi$ 
while it is $2\pi$ for MI-III and
MI-IV identities. Further, whereas the MI-I and MI-II identities are valid
for odd as well as even $p$, for MI-III and MI-IV cases, nontrivial identities
are obtained only when $p$ is an odd integer. 
It is worth keeping in mind that the identities given in this paper are not 
exhaustive but are meant to be representative identities of low rank.   

\section{Identities following from MI-I.}

As shown in IIa, one of the simplest MI-I identities is given by (see
Eq. (22)
of IIa)
\beq\label{3}
\sum_{j=1}^p \dn(x_j)\dn(x_{j+1})\dn(x_{j+2}) \,=\,
\left[\cs^2(2K/p)-2\cs(2K/p)\cs(4K/p)\right] \sum_{j=1}^{p} \dn(x_j)~,
\eeq
where
\beq\label{4}
x_j = u +(j-1)T/p~,
\eeq
with $T$ being $2K$ or $4K$ depending on whether the identity is of type MI-I,II  or
type MI-III,IV  respectively. 
On using the relation between Jacobi theta and Jacobi elliptic 
functions [Eqs. (\ref{1}) and (\ref{2})], the corresponding identity
in terms of theta functions is
\beq\label{5}
\sum_{j=1}^{p} \frac{\theta_3 (z_j) \theta_3 (z_{j+1}) \theta_3 (z_{j+2})}
{\theta_4 (z_{j}) \theta_4 (z_{j+1}) \theta_4 (z_{j+2})} = \frac{\theta_2 (\pi /p)}
{\theta_1 (\pi /p)}\left[\frac{\theta_2 (\pi /p)}{\theta_1 (\pi /p)}
-2\frac{\theta_2 (2\pi /p)}{\theta_1 (2\pi /p)}\right]\sum_{j=1}^{p} 
\frac{\theta_3 (z_j)}{\theta_4 (z_j)}~.
\eeq
Here $z_j \equiv z+(j-1)\pi /p$ with $z = u \pi /2K = u /\theta_3^2 (0)$.

Proceeding in the same way we now rewrite the various MI-I identities obtained
in IIa in terms of the ratios of Jacobi theta functions. 
For example  
identities (83) to (111) [except identities (105) and (109)] 
of IIa take the forms given below:
\beq\label{7}
\sum_{j=1}^{p} \frac{\theta_1 (z_j)}{\theta_4 (z_j)}\left[
\frac{\theta_2 (z_{j+r})}{\theta_4 (z_{j+r})}+
\frac{\theta_2 (z_{j-r})}{\theta_4 (z_{j-r})}\right] = 0~.
\eeq
\beqa\label{8}
&&\sum_{j=1}^{p} \frac{\theta_3 (z_j)}{\theta_4 (z_j)}\frac{\theta_3 (z_{j+r})}
{\theta_4 (z_{j+r})}...\frac{\theta_3 (z_{j+(l-1)r})}{\theta_4 (z_{j+(l-1)r})} 
=\bigg [\Pi_{k=1}^{\frac{(l-1)}{2}} \frac{\theta_2^2 (rk\pi /p)}
{\theta_1^2 (rk\pi /p)} \nonumber \\
&&+ 2(-1)^{\frac{l-1}{2}} \left(\frac{\theta_4 (0)}
{\theta_3 (0)}\right)^{\frac{(l-1)(l-3)}{2}}
\sum_{k=1}^{(l-1)/2} \Pi_{n\ne k,n =1}^{l} \frac{\theta_2 ([n-k]r\pi /p)}
{\theta_1 ([n-k]r \pi /p)} \bigg ] 
\sum_{j=1}^{p} \frac{\theta_3 (z_j)}{\theta_4 (z_j)}~,
\eeqa
where $l$ is any odd integer ($\ge 3$).
In the special case when $p$ is also odd and $l=p$, this identity takes the 
elegant form
\beq\label{9}
\Pi_{j=1}^{p} \frac{\theta_3 (z_j)}{\theta_4 (z_j)} = 
\Pi_{n=1}^{\frac{(p-1)}{2}}
\frac{\theta_2^2 (n\pi /p)}{\theta_1^2 (n\pi /p)} \sum_{j=1}^{p} 
\frac{\theta_3 (z_j)}{\theta_4 (z_j)}~.
\eeq
\beqa\label{10}
&&\sum_{j=1}^{p} \frac{\theta_3^2 (z_j)}{\theta_4^2 (z_{j})}
\left[\frac{\theta_3 (z_{j+r})}{\theta_4 (z_{j+r})} +
\frac{\theta_3 (z_{j-r})}{\theta_4 (z_{j-r})}\right] \nonumber \\ 
&&= 2\left[\frac{\theta_2^2 (0)}
{\theta_3 (0) \theta_4 (0)} \frac{\theta_3 (r\pi /p) \theta_4 (r\pi /p)}
{\theta_1^2 (r\pi /p)}-\frac{\theta_2^2 (r\pi /p)}{\theta_1^2 (r\pi /p)}\right] 
\sum_{j=1}^{p} 
\frac{\theta_3 (z_j)}{\theta_4 (z_j)}~.
\eeqa
\beqa\label{11}
&&\sum_{j=1}^{p} \frac{\theta_2 (z_j)}{\theta_4 (z_{j})}
\left[\frac{\theta_2 (z_{j+r}) \theta_3 (z_{j+r})}{\theta_4^2 (z_{j+r})} +
\frac{\theta_2 (z_{j-r}) \theta_3 (z_{j-r})}{\theta_4^2 (z_{j-r})}\right] \nonumber \\ 
&&= -2\frac{\theta_3 (0)}
{\theta_2 (0)} \frac{\theta_2 (r\pi /p)}{\theta_1 (r\pi /p)} 
\left[\frac{\theta_3 (r\pi /p)}{\theta_1 (r\pi /p)}
-\frac{\theta_3 (0)}{\theta_4 (0)}
\frac{\theta_4 (r\pi /p)}{\theta_1 (r\pi /p)}\right] 
\sum_{j=1}^{p} 
\frac{\theta_3 (z_j)}{\theta_4 (z_j)}~.
\eeqa
\beqa\label{12}
&&\sum_{j=1}^{p} \frac{\theta_1 (z_j)}{\theta_4 (z_{j})}
\left[\frac{\theta_1 (z_{j+r}) \theta_3 (z_{j+r})}{\theta_4^2 (z_{j+r})} +
\frac{\theta_1 (z_{j-r}) \theta_3 (z_{j-r})}{\theta_4^2 (z_{j-r})}\right] \nonumber \\ 
&&= -2\frac{\theta_4^2 (0)}
{\theta_2 (0) \theta_3 (0)} \frac{\theta_2 (r\pi /p)}{\theta_1 (r\pi /p)} 
\left[\frac{\theta_3 (r\pi /p)}{\theta_1 (r\pi /p)}
-\frac{\theta_3 (0)}{\theta_4 (0)}
\frac{\theta_4 (r\pi /p)}{\theta_1 (r\pi /p)}\right] 
\sum_{j=1}^{p} 
\frac{\theta_3 (z_j)}{\theta_4 (z_j)}~.
\eeqa
\beqa\label{13}
&&\sum_{j=1}^{p} \frac{\theta_3 (z_j)}{\theta_4 (z_{j})}
\left[\frac{\theta_3 (z_{j+r}) \theta_3 (z_{j+s})}
{\theta_4 (z_{j+r}) \theta_4 (z_{j+s})} +
\frac{\theta_3 (z_{j-r}) \theta_3 (z_{j-s})}
{\theta_4 (z_{j-r}) \theta_4 (z_{j-s})}\right] \nonumber \\ 
&&= -2
\left[\frac{\theta_2 (r\pi /p) \theta_2 (s\pi /p)}
{\theta_1 (r\pi /p) \theta_1 (s\pi /p)}
+\frac{\theta_2 ([r-s]\pi /p)}{\theta_1 ([r-s]\pi /p)}
\left(\frac{\theta_2 (r\pi /p)}{\theta_1 (r\pi /p)}
-\frac{\theta_2 (s\pi /p)}{\theta_1 (s\pi /p)}\right)
\right] 
\sum_{j=1}^{p} 
\frac{\theta_3 (z_j)}{\theta_4 (z_j)}~.
\eeqa
\beqa\label{14}
&&\sum_{j=1}^{p} \frac{\theta_3 (z_j)}{\theta_4 (z_{j})}
\left[\frac{\theta_2 (z_{j+r}) \theta_2 (z_{j+s})}
{\theta_4 (z_{j+r}) \theta_4 (z_{j+s})} +
\frac{\theta_2 (z_{j-r}) \theta_2 (z_{j-s})}
{\theta_4 (z_{j-r}) \theta_4 (z_{j-s})}\right] \nonumber \\ 
&&= -2
\left[\frac{\theta_3 (r\pi /p) \theta_3 (s\pi /p)}
{\theta_1 (r\pi /p) \theta_1 (s\pi /p)}
+\frac{\theta_3 ([r-s]\pi /p) \theta_3 (0)}{\theta_1 ([r-s]\pi /p) \theta_2 (0)}
\left(\frac{\theta_2 (r\pi /p)}{\theta_1 (r\pi /p)}
-\frac{\theta_2 (s\pi /p)}{\theta_1 (s\pi /p)}\right)
\right] 
\sum_{j=1}^{p} 
\frac{\theta_3 (z_j)}{\theta_4 (z_j)}~.
\eeqa
\beqa\label{15}
&&\sum_{j=1}^{p} \frac{\theta_3 (z_j)}{\theta_4 (z_{j})}
\left[\frac{\theta_1 (z_{j+r}) \theta_1 (z_{j+s})}
{\theta_4 (z_{j+r}) \theta_4 (z_{j+s})} +
\frac{\theta_1 (z_{j-r}) \theta_1 (z_{j-s})}
{\theta_4 (z_{j-r}) \theta_4 (z_{j-s})}\right] \nonumber \\ 
&&= 2
\left[\frac{\theta_4 (r\pi /p) \theta_4 (s\pi /p)}
{\theta_1 (r\pi /p) \theta_1 (s\pi /p)}
+\frac{\theta_4 ([r-s]\pi /p) \theta_4 (0)}{\theta_1 ([r-s]\pi /p) \theta_2 (0)}
\left(\frac{\theta_2 (r\pi /p)}{\theta_1 (r\pi /p)}
-\frac{\theta_2 (s\pi /p)}{\theta_1 (s\pi /p)}\right)
\right] 
\sum_{j=1}^{p} 
\frac{\theta_3 (z_j)}{\theta_4 (z_j)}~.
\eeqa
\beqa\label{16}
&&\sum_{j=1}^{p} \frac{\theta_2 (z_j)}{\theta_4 (z_{j})}
\left[\frac{\theta_3 (z_{j+r}) \theta_2 (z_{j+s})}
{\theta_4 (z_{j+r}) \theta_4 (z_{j+s})} +
\frac{\theta_3 (z_{j-r}) \theta_2 (z_{j-s})}
{\theta_4 (z_{j-r}) \theta_4 (z_{j-s})}\right]  
= -2
\bigg[\frac{\theta_3 (r\pi /p) \theta_3 (0)}
{\theta_1 (r\pi /p) \theta_2 (0)} \nonumber \\
&&\left(\frac{\theta_2 (s\pi /p)}
{\theta_1 (s\pi/p)}+\frac{\theta_2 ([r-s]\pi /p)}
{\theta_1 ([r-s]\pi /p)}\right)-
\frac{\theta_3 ([r-s]\pi /p) \theta_3 (s\pi /p)}
{\theta_1 ([r-s]\pi /p) \theta_1 (s\pi /p)}\bigg]
\sum_{j=1}^{p} 
\frac{\theta_3 (z_j)}{\theta_4 (z_j)}~.
\eeqa
\beqa\label{17}
&&\sum_{j=1}^{p} \frac{\theta_1 (z_j)}{\theta_4 (z_{j})}
\left[\frac{\theta_3 (z_{j+r}) \theta_1 (z_{j+s})}
{\theta_4 (z_{j+r}) \theta_4 (z_{j+s})} +
\frac{\theta_3 (z_{j-r}) \theta_1 (z_{j-s})}
{\theta_4 (z_{j-r}) \theta_4 (z_{j-s})}\right]  
= 2
\bigg[\frac{\theta_4 (r\pi /p) \theta_4 (0)}
{\theta_1 (r\pi /p) \theta_2 (0)} \nonumber \\
&&\left(\frac{\theta_2 (s\pi /p)}
{\theta_1 (s\pi/p)}+\frac{\theta_2 ([r-s]\pi /p)}
{\theta_1 ([r-s]\pi /p)}\right)-
\frac{\theta_4 ([r-s]\pi /p) \theta_4 (s\pi /p)}
{\theta_1 ([r-s]\pi /p) \theta_1 (s\pi /p)}\bigg]
\sum_{j=1}^{p} 
\frac{\theta_3 (z_j)}{\theta_4 (z_j)}~.
\eeqa
\beqa\label{18}
&&\sum_{j=1}^{p} \frac{\theta_1^2 (z_j)}{\theta_4^2 (z_{j})}
\left[\frac{\theta_2 (z_{j+r}) \theta_2 (z_{j+s}) \theta_3 (z_{j+t})}
{\theta_4 (z_{j+r}) \theta_4 (z_{j+s}) \theta_4 (z_{j+t})} +
\frac{\theta_2 (z_{j-r}) \theta_2 (z_{j-s}) \theta_3 (z_{j-t})}
{\theta_4 (z_{j-r}) \theta_4 (z_{j-s}) \theta_4 (z_{j-t})}\right] \nonumber \\ 
&&= 2
\bigg[\frac{\theta_3 (0) \theta_4 (0)}{\theta_2^2 (0)} 
\left(\frac{\theta_2 (r\pi /p) \theta_2 (t\pi /p) \theta_3 (s\pi /p) 
\theta_4 (r\pi /p)}
{\theta_1^2 (r\pi /p) \theta_1 (t\pi /p) \theta_1 (s\pi /p)}+
\frac{\theta_2 (s\pi /p) \theta_2 (t\pi /p) \theta_3 (r\pi /p) 
\theta_4 (s\pi /p)}
{\theta_1^2 (s\pi /p) \theta_1 (t\pi /p) \theta_1 (r\pi /p)}\right) 
\nonumber \\ 
&&+\frac{\theta_4 (0)}{\theta_3 (0)}
\frac{\theta_3 (r\pi /p) \theta_3 (s\pi /p) \theta_3 (t\pi /p) 
\theta_4 (t\pi /p)}
{\theta_1^2 (t\pi /p) \theta_1 (r\pi /p) \theta_1 (s\pi /p)}  
-\frac{\theta_3 ([r-t]\pi /p) \theta_3 ([s-t]\pi /p) \theta_4^2 (s\pi /p)} 
{\theta_1^2 (s\pi /p) \theta_1 ([r-t]\pi /p) \theta_1 ([s-t]\pi /p)} \nonumber \\
&&-\frac{\theta_3 (0)}{\theta_2 (0)}
\bigg(\frac{\theta_2 ([r-t]\pi /p) \theta_3 ([r-s]\pi /p) \theta_4^2 (r\pi /p)} 
{\theta_1^2 (r\pi /p) \theta_1 ([r-t]\pi /p) \theta_1 ([r-s]\pi /p)} \nonumber \\ 
&&+\frac{\theta_2 ([t-s]\pi /p) \theta_3 ([r-s]\pi /p) \theta_4^2 (s\pi /p)} 
{\theta_1^2 (s\pi /p) \theta_1 ([t-s]\pi /p) \theta_1 ([r-s]\pi /p)}\bigg)
\bigg ] \sum_{j=1}^{p} \frac{\theta_3 (z_j)}{\theta_4 (z_j)}~.
\eeqa
\beqa\label{19}
&&\sum_{j=1}^{p} \frac{\theta_3^2 (z_j)}{\theta_4^2 (z_{j})}
\left[\frac{\theta_2 (z_{j+r}) \theta_1 (z_{j+r})}
{\theta_4^2 (z_{j+r})} +
\frac{\theta_2 (z_{j-r}) \theta_1 (z_{j-r})}
{\theta_4^2 (z_{j-r})}\right] \nonumber \\ 
&&= -2 \frac{\theta_2^2 (r\pi /p)}{\theta_1^2 (r\pi /p)}
\left[1+\frac{\theta_2^2 (0) \theta_3 (r\pi /p) \theta_4 (r\pi /p)}
{\theta_2^2 (r\pi /p) \theta_3 (0) \theta_4 (0)}\right]
\sum_{j=1}^{p} \frac{\theta_1 (z_j) \theta_2 (z_j)}{\theta_4^2 (z_j)}~.
\eeqa
\beq\label{6}
\sum_{j=1}^{p} \frac{\theta_1 (z_j) \theta_2 (z_{j}) \theta_3 (z_{j})}
{\theta_4^3 (z_{j})} \left[\frac{\theta_3 (z_{j+r})}{\theta_4 (z_{j+r})} +
\frac{\theta_3 (z_{j-r})}{\theta_4 (z_{j-r})}\right]=  
\frac{\theta_2^2 (0) \theta_3 (r\pi /p) \theta_4 (r\pi /p)}
{\theta_3 (0) \theta_4 (0) \theta_1^2 (r\pi /p)} \sum_{j=1}^{p} 
\frac{\theta_1 (z_j) \theta_2 (z_{j})}{\theta_4^2 (z_j)}~.
\eeq
\beqa\label{20}
&&\sum_{j=1}^{p} \frac{\theta_1 (z_j) \theta_3 (z_j)}{\theta_4^2 (z_{j})}
\left[\frac{\theta_2 (z_{j+r}) \theta_3 (z_{j+r})}
{\theta_4^2 (z_{j+r})} +
\frac{\theta_2 (z_{j-r}) \theta_3 (z_{j-r})}
{\theta_4^2 (z_{j-r})}\right] \nonumber \\ 
&&= -2 \frac{\theta_2 (0) \theta_2 (r\pi /p)}{\theta_1^2 (r\pi /p)}
\left[\frac{\theta_4 (r\pi /p)}{\theta_4 (0)}
+\frac{\theta_3 (r\pi /p)}{\theta_3 (0)}\right]
\sum_{j=1}^{p} \frac{\theta_1 (z_j) \theta_2 (z_j)}{\theta_4^2 (z_j)}~.
\eeqa
\beqa\label{20a}
&&\sum_{j=1}^{p} \frac{\theta_2 (z_j)}{\theta_4 (z_{j})}
\left[\frac{\theta_1^3 (z_{j+r})}
{\theta_4^3 (z_{j+r})} +
\frac{\theta_1^3 (z_{j-r})}
{\theta_4^3 (z_{j-r})}\right] \nonumber \\ 
&&= -2 \frac{\theta_4 (0) \theta_2 (r\pi /p)\theta_4 (r\pi /p)}
{\theta_2 (0) \theta_1^2 (r\pi /p)}
\sum_{j=1}^{p} \frac{\theta_1 (z_j) \theta_2 (z_j)}{\theta_4^2 (z_j)}~.
\eeqa
\beqa\label{20b}
&&\sum_{j=1}^{p} \frac{\theta_1 (z_j)}{\theta_4 (z_{j})}
\left[\frac{\theta_2^3 (z_{j+r})}
{\theta_4^3 (z_{j+r})} +
\frac{\theta_2^3 (z_{j-r})}
{\theta_4^3 (z_{j-r})}\right] \nonumber \\ 
&&= 2 \frac{\theta_3 (0) \theta_2 (r\pi /p)\theta_3 (r\pi /p)}
{\theta_2 (0) \theta_1^2 (r\pi /p)}
\sum_{j=1}^{p} \frac{\theta_1 (z_j) \theta_2 (z_j)}{\theta_4^2 (z_j)}~.
\eeqa
\beqa\label{21}
&&\sum_{j=1}^{p} \frac{\theta_1 (z_j) \theta_2 (z_j)}{\theta_4^2 (z_{j})}
\left[\frac{\theta_3 (z_{j+r}) \theta_3 (z_{j+s})}
{\theta_4 (z_{j+r}) \theta_4 (z_{j+s})} +
\frac{\theta_3 (z_{j-r}) \theta_3 (z_{j-s})}
{\theta_4 (z_{j-r}) \theta_4 (z_{j-s})}\right] \nonumber \\
&&= -2 \frac{\theta_2 (r\pi /p) \theta_2 (s\pi /p)}
{\theta_1 (r\pi /p) \theta_1 (s\pi /p)}
\sum_{j=1}^{p} \frac{\theta_1 (z_j) \theta_2 (z_j)}{\theta_4^2 (z_j)}~.
\eeqa
\beqa\label{22}
&&\sum_{j=1}^{p} \frac{\theta_1 (z_j) \theta_2 (z_j)}{\theta_4^2 (z_{j})}
\left[\frac{\theta_2 (z_{j+r}) \theta_2 (z_{j+s})}
{\theta_4 (z_{j+r}) \theta_4 (z_{j+s})} +
\frac{\theta_2 (z_{j-r}) \theta_2 (z_{j-s})}
{\theta_4 (z_{j-r}) \theta_4 (z_{j-s})}\right] \nonumber \\
&&= -2 \frac{\theta_3 (r\pi /p) \theta_3 (s\pi /p)}
{\theta_1 (r\pi /p) \theta_1 (s\pi /p)}
\sum_{j=1}^{p} \frac{\theta_1 (z_j) \theta_2 (z_j)}{\theta_4^2 (z_j)}~.
\eeqa
\beqa\label{23}
&&\sum_{j=1}^{p} \frac{\theta_1 (z_j) \theta_2 (z_j)}{\theta_4^2 (z_{j})}
\left[\frac{\theta_1 (z_{j+r}) \theta_1 (z_{j+s})}
{\theta_4 (z_{j+r}) \theta_4 (z_{j+s})} +
\frac{\theta_1 (z_{j-r}) \theta_1 (z_{j-s})}
{\theta_4 (z_{j-r}) \theta_4 (z_{j-s})}\right] \nonumber \\
&&= 2 \frac{\theta_4 (r\pi /p) \theta_4 (s\pi /p)}
{\theta_1 (r\pi /p) \theta_1 (s\pi /p)}
\sum_{j=1}^{p} \frac{\theta_1 (z_j) \theta_2 (z_j)}{\theta_4^2 (z_j)}~.
\eeqa
\beqa\label{24}
&&\sum_{j=1}^{p} \frac{\theta_2 (z_j) \theta_3 (z_j)}{\theta_4^2 (z_{j})}
\left[\frac{\theta_1 (z_{j+r}) \theta_3 (z_{j+s})}
{\theta_4 (z_{j+r}) \theta_4 (z_{j+s})} +
\frac{\theta_1 (z_{j-r}) \theta_3 (z_{j-s})}
{\theta_4 (z_{j-r}) \theta_4 (z_{j-s})}\right] \nonumber \\ 
&&= -2 \frac{\theta_2 (0) \theta_4 (r\pi /p) \theta_2 (s\pi /p)}
{\theta_4 (0) \theta_1 (r\pi /p) \theta_1 (s\pi /p)}
\sum_{j=1}^{p} \frac{\theta_1 (z_j) \theta_2 (z_j)}{\theta_4^2 (z_j)}~.
\eeqa
\beqa\label{24a}
&&\sum_{j=1}^{p} \frac{\theta_1 (z_j) \theta_3 (z_j)}{\theta_4^2 (z_{j})}
\left[\frac{\theta_2 (z_{j+r}) \theta_3 (z_{j+s})}
{\theta_4 (z_{j+r}) \theta_4 (z_{j+s})} +
\frac{\theta_2 (z_{j-r}) \theta_3 (z_{j-s})}
{\theta_4 (z_{j-r}) \theta_4 (z_{j-s})}\right] \nonumber \\
&&= -2 \frac{\theta_2 (0) \theta_3 (r\pi /p) \theta_2 (s\pi /p)}
{\theta_3 (0) \theta_1 (r\pi /p) \theta_1 (s\pi /p)}
\sum_{j=1}^{p} \frac{\theta_1 (z_j) \theta_2 (z_j)}{\theta_4^2 (z_j)}~.
\eeqa
\beqa\label{25}
&&\sum_{j=1}^{p} \frac{\theta_1 (z_j) \theta_2 (z_j) \theta_3 (z_j)}
{\theta_4^3 (z_{j})}
\left[\frac{\theta_3^3 (z_{j+r})}
{\theta_4^3 (z_{j+r})} +
\frac{\theta_3^3 (z_{j-r})}
{\theta_4^3 (z_{j-r})}\right]  
= - \frac{2\theta_2^2 (0) \theta_2^2 (r\pi /p)}{3\theta_1^4(r\pi /p)}
\bigg[\frac{\theta_4^2 (r\pi /p)}{\theta_4^2 (0)} \nonumber \\
&&+\frac{\theta_2^2 (0) \theta_3^2 (r\pi /p) \theta_4^2 (r\pi /p)}
{\theta_2^2 (r\pi /p) \theta_3^2 (0) \theta_4^2 (0)}+
\frac{\theta_3^2 (r\pi /p)}{\theta_3^2 (0)} 
+3\frac{\theta_3 (r\pi /p) \theta_4 (r\pi /p)}
{\theta_3 (0) \theta_4 (0)}\bigg] 
\sum_{j=1}^{p} \frac{\theta_1 (z_j) \theta_2 (z_j)}{\theta_4^2 (z_j)}~.
\eeqa
\beqa\label{26}
&&\sum_{j=1}^{p} \frac{\theta_3^4 (z_j)}
{\theta_4^4 (z_{j})}\left[\frac{\theta_3 (z_{j+r})}
{\theta_4 (z_{j+r})} +
\frac{\theta_3 (z_{j-r})}
{\theta_4 (z_{j-r})}\right]  
= 2 \frac{\theta_2^2 (0) \theta_3 (r\pi /p) \theta_4 (r\pi /p)}
{\theta_3 (0) \theta_4 (0) \theta_1^2 (r\pi /p)}
\sum_{j=1}^{p} \frac{\theta_3^3 (z_j)}{\theta_4^3 (z_j)} \nonumber \\
&&+2\frac{\theta_2^4 (r\pi /p)}{\theta_1^4 (r\pi /p)}
\left[1-
\frac{\theta_2^2 (0) \theta_3 (r\pi /p) \theta_4 (r\pi /p)}
{\theta_2^2 (r\pi /p) \theta_3 (0) \theta_4 (0)}\right]
\sum_{j=1}^{p} \frac{\theta_3 (z_j)}{\theta_4 (z_j)}~.
\eeqa
\beqa\label{27}
&&\sum_{j=1}^{p} \frac{\theta_3^3 (z_j)}
{\theta_4^3 (z_{j})}\left[\frac{\theta_3^2 (z_{j+r})}
{\theta_4^2 (z_{j+r})} +
\frac{\theta_3^2 (z_{j-r})}
{\theta_4^2 (z_{j-r})}\right]  
= -2 \frac{\theta_2^2 (r\pi /p)}
{\theta_1^2 (r\pi /p)}
\sum_{j=1}^{p} \frac{\theta_3^3 (z_j)}{\theta_4^3 (z_j)} \nonumber \\
&&+2\frac{\theta_2^2 (0) \theta_2^2 (r\pi /p)}{\theta_1^4 (r\pi /p)}
\bigg[
\frac{\theta_4^2 (r\pi /p)}{\theta_4^2 (0)}+
\frac{\theta_3^2 (r\pi /p)}{\theta_3^2 (0)} \nonumber \\
&&+\frac{\theta_2^2 (0) \theta_3^2 (r\pi /p) \theta_4^2 (r\pi /p)}
{\theta_2^2 (r\pi /p) \theta_3^2 (0) \theta_4^2 (0)}
-3\frac{\theta_3 (r\pi /p) \theta_4 (r\pi /p)}
{\theta_3 (0) \theta_4 (0)}\bigg]
\sum_{j=1}^{p} \frac{\theta_3 (z_j)}{\theta_4 (z_j)}~.
\eeqa
\beqa\label{28}
&&\sum_{j=1}^{p} \frac{\theta_3^4 (z_j)}
{\theta_4^4 (z_{j})}\left[\frac{\theta_1 (z_{j+r}) \theta_2 (z_{j+r})}
{\theta_4^2 (z_{j+r})} +
\frac{\theta_1 (z_{j-r}) \theta_2 (z_{j-r})}
{\theta_4^2 (z_{j-r})}\right] \nonumber \\ 
&&= -2 \frac{\theta_2^2 (0) \theta_3 (r\pi /p) \theta_4 (r\pi /p)}
{\theta_3 (0) \theta_4 (0) \theta_1^2 (r\pi /p)}
\sum_{j=1}^{p} \frac{\theta_1 (z_{j}) \theta_2 (z_j) \theta_3^2 (z_j)}
{\theta_4^4 (z_j)} \nonumber \\
&&+2\frac{\theta_2^4 (r\pi /p)}{\theta_1^4 (r\pi /p)}
\left[1+
3\frac{\theta_2^2 (0) \theta_3 (r\pi /p) \theta_4 (r\pi /p)}
{\theta_2^2 (r\pi /p) \theta_3 (0) \theta_4 (0)}\right]
\sum_{j=1}^{p} \frac{\theta_1 (z_j) \theta_2 (z_j)}{\theta_4^2 (z_j)}~.
\eeqa
\beqa\label{29}
&&\sum_{j=1}^{p} \frac{\theta_3^4 (z_j)}
{\theta_4^4 (z_{j})}\left[\frac{\theta_1 (z_{j+r}) \theta_2 (z_{j+s})}
{\theta_4 (z_{j+r}) \theta_4 (z_{j+s}} +
\frac{\theta_1 (z_{j-r}) \theta_2 (z_{j-s})}
{\theta_4 (z_{j-r}) \theta_4 (z_{j-s}}\right]  
= -2 \frac{\theta_2^2 (0) \theta_4 (r\pi /p) \theta_3 (s\pi /p)}
{\theta_3 (0) \theta_4 (0) \theta_1 (r\pi /p) \theta_1 (s\pi /p)} \nonumber \\
&&\sum_{j=1}^{p} \frac{\theta_1 (z_{j}) \theta_2 (z_j) \theta_3^2 (z_j)}
{\theta_4^4 (z_j)} 
+2\frac{\theta_2^2 (0)}{\theta_3 (0) \theta_4 (0)}
\bigg[
\frac{\theta_2 (r\pi /p) \theta_2 (s\pi /p) \theta_3 (r\pi /p) \theta_4 (s\pi /p)}
{\theta_1^2 (r\pi /p) \theta_1^2 (s\pi /p)} \nonumber \\
&&+\frac{\theta_3 (s\pi /p) \theta_4 (r\pi /p)}
{\theta_1 (r\pi /p) \theta_1 (s\pi /p)} 
\left(\frac{\theta_2^2 (r\pi /p)}{\theta_1^2 (r\pi /p)}+
\frac{\theta_2^2 (s\pi /p)}{\theta_1^2 (s\pi /p)}\right)\bigg]
\sum_{j=1}^{p} \frac{\theta_1 (z_j) \theta_2 (z_j)}{\theta_4^2 (z_j)}~.
\eeqa

\sss

\subsection{MI-I Identities with alternating signs.}

Let us now write a few MI-I identities with alternate signs in terms of the 
ratios of theta functions. As emphasized in II, such identities are only valid
when $p$ is even and hence $r$ (being co-prime to $p$) is necessarily odd. 

 For example, the identities  
(196) to (201), (207), (208) and (215)  of IIa take the form given below:
\beq\label{32}
\sum_{j=1}^{p} (-1)^{j-1} \frac{\theta_1 (z_j)}
{\theta_4 (z_{j})}\left[\frac{\theta_2 (z_{j+1})}
{\theta_4 (z_{j+1})} +
\frac{\theta_2 (z_{j-1})}
{\theta_4 (z_{j-1})}\right] =0~. 
\eeq
\beqa\label{34}
&&\sum_{j=1}^{p} (-1)^{j-1} \frac{\theta_3 (z_j) \theta_3 (z_{j+r}) 
\theta_3 (z_{j+2r})}
{\theta_4 (z_{j}) \theta_4 (z_{j+r}) \theta_4 (z_{j+2r})} \nonumber \\
&&=-\left[\frac{\theta_2^2 (r\pi /p)}{\theta_1^2 (r\pi /p)}+
\frac{2\theta_2 (r\pi /p) \theta_2 (2r\pi /p)}
{\theta_1 (r\pi /p) \theta_1 (2r\pi /p)}\right]
\sum_{j=1}^{p} (-1)^{j-1} \frac{\theta_3 (z_j)}
{\theta_4 (z_j)}~.
\eeqa
\beqa\label{35}
&&\sum_{j=1}^{p} (-1)^{j-1} \frac{\theta_3 (z_j) \theta_3 (z_{j+r}) 
\theta_3 (z_{j+s})}
{\theta_4 (z_{j}) \theta_4 (z_{j+r}) \theta_4 (z_{j+s})}
=-\bigg[\frac{\theta_2 (r\pi /p) \theta_2 (s\pi /p)}
{\theta_1 (r\pi /p) \theta_1 (s\pi /p)} \nonumber \\
&&-\frac{\theta_2 (r\pi /p) \theta_2 ([r-s]\pi /p)}
{\theta_1 (r\pi /p) \theta_1 ([r-s]\pi /p)}-
\frac{\theta_2 (s\pi /p) \theta_2 ([s-r]\pi /p)}
{\theta_1 (s\pi /p) \theta_1 ([s-r]\pi /p)}\bigg]
\sum_{j=1}^{p} (-1)^{j-1} \frac{\theta_3 (z_j)}
{\theta_4 (z_j)}~.
\eeqa
\beqa\label{31}
&&\sum_{j=1}^{p} (-1)^{j-1} \frac{\theta_3^2 (z_j)}
{\theta_4^2 (z_{j})}\left[\frac{\theta_3 (z_{j+r})}
{\theta_4 (z_{j+r})} +
\frac{\theta_3 (z_{j-r})}
{\theta_4 (z_{j-r})}\right] \nonumber \\  
&&=2\frac{\theta_2^2 (0)}{\theta_1^2 (r\pi /p)}
\bigg[\frac{\theta_3 (r\pi /p) \theta_4 (r\pi /p)}
{\theta_3 (0) \theta_4 (0)}
+ \frac{\theta_2^2 (r\pi /p)}
{\theta_2^2 (0)}\bigg]
\sum_{j=1}^{p} (-1)^{j-1} \frac{\theta_3 (z_j)}{\theta_4 (z_j)}~.
\eeqa
\beqa\label{36a}
&&\sum_{j=1}^{p} (-1)^{j-1} \frac{\theta_2 (z_j)}
{\theta_4 (z_{j})}\left[\frac{\theta_2 (z_{j+r}) \theta_3 (z_{j+r})}
{\theta_4^2 (z_{j+r})} +
\frac{\theta_2 (z_{j-r}) \theta_3 (z_{j-r})}
{\theta_4^2 (z_{j-r})}\right] \nonumber \\ 
&&=-2\frac{\theta_3^2 (0) \theta_2 (r\pi /p)}
{\theta_2 (0) \theta_1^2 (r\pi /p)}
\left[\frac{\theta_3 (r\pi /p)}
{\theta_3 (0)} 
+\frac{\theta_4 (r\pi /p)}
{\theta_4 (0)}\right]
\sum_{j=1}^{p} (-1)^{j-1} \frac{\theta_3 (z_j)}{\theta_4 (z_j)}~.
\eeqa
\beqa\label{37}
&&\sum_{j=1}^{p} (-1)^{j-1} \frac{\theta_1 (z_j)}
{\theta_4 (z_{j})}\left[\frac{\theta_1 (z_{j+r}) \theta_3 (z_{j+r})}
{\theta_4^2 (z_{j+r})} +
\frac{\theta_1 (z_{j-r}) \theta_3 (z_{j-r})}
{\theta_4^2 (z_{j-r})}\right] \nonumber \\ 
&&=2\frac{\theta_3^2 (0) \theta_2 (r\pi /p)}
{\theta_2 (0) \theta_1^2 (r\pi /p)}
\left[\frac{\theta_3 (r\pi /p)}
{\theta_3 (0)} 
+\frac{\theta_4 (r\pi /p)}
{\theta_4 (0)}\right]
\sum_{j=1}^{p} (-1)^{j-1} \frac{\theta_3 (z_j)}{\theta_4 (z_j)}~.
\eeqa
\beqa\label{38}
&&\sum_{j=1}^{p} (-1)^{j-1} \frac{\theta_3^2 (z_j)}
{\theta_4^2 (z_{j})}\left[\frac{\theta_1 (z_{j+r}) \theta_2 (z_{j+r})}
{\theta_4^2 (z_{j+r})} +
\frac{\theta_1 (z_{j-r}) \theta_2 (z_{j-r})}
{\theta_4^2 (z_{j-r})}\right] \nonumber \\ 
&&=2\frac{\theta_2^2 (0)}{\theta_1^2 (r\pi /p)}
\left[\frac{\theta_2^2 (r\pi /p)}
{\theta_2^2 (0)}
-\frac{\theta_3 (r\pi /p) \theta_4 (r\pi /p)}
{\theta_3 (0) \theta_4 (0)}\right]
\sum_{j=1}^{p} (-1)^{j-1} 
\frac{\theta_1 (z_j) \theta_2 (z_j)}{\theta_4^2 (z_j)}~.
\eeqa
\beqa\label{38a}
&&\sum_{j=1}^{p} (-1)^{j-1} \frac{\theta_1 (z_j) \theta_3 (z_j)}
{\theta_4^2 (z_{j})}\left[\frac{\theta_2 (z_{j+r}) \theta_3 (z_{j+r})}
{\theta_4^2 (z_{j+r})} +
\frac{\theta_2 (z_{j-r}) \theta_3 (z_{j-r})}
{\theta_4^2 (z_{j-r})}\right] \nonumber \\ 
&&=-2\frac{\theta_2 (0) \theta_2 (r\pi /p)}{\theta_1^2 (r\pi /p)}
\left[\frac{\theta_3 (r\pi /p)}
{\theta_3 (0)}
+\frac{\theta_4 (r\pi /p)}
{\theta_4 (0)}\right]
\sum_{j=1}^{p} (-1)^{j-1} 
\frac{\theta_1 (z_j) \theta_2 (z_j)}{\theta_4^2 (z_j)}~.
\eeqa
\beqa\label{33}
&&\sum_{j=1}^{p} (-1)^{j-1} \frac{\theta_1 (z_j) \theta_2 (z_j) 
\theta_3 (z_j)}
{\theta_4^3 (z_{j})}\left[\frac{\theta_3 (z_{j+1})}
{\theta_4 (z_{j+1})} +
\frac{\theta_3 (z_{j-1})}
{\theta_4 (z_{j-1})}\right] \nonumber \\  
&&=2\frac{\theta_2^2 (0) \theta_3 (\pi /p) \theta_4 (\pi /p)}
{\theta_3 (0) \theta_4 (0) \theta_1^2 (\pi /p)}
\sum_{j=1}^{p} (-1)^{j-1} \frac{\theta_1 (z_j) \theta_2 (z_j)}
{\theta_4^2 (z_j)}~.
\eeqa

\section{Identities following from MI-II.}

We shall now write down some of the MI-II identities from IIa in terms
of the ratios of theta functions. As in the previous section, in this section too 
$z_j \equiv z+(j-1)\pi /p$ with $z = u \pi /2K = u /\theta_3^2 (0)$.
Identities (112) to (141) 
[except identities (120), (121), (126), (130) to (133) and (136)] of IIa, when expressed in terms of theta functions, are given by
\beq\label{39}
\sum_{j=1}^{p} \frac{\theta_3 (z_j) \theta_3 (z_{j+r})}
{\theta_4 (z_j) \theta_4 (z_{j+r})} 
=\frac{p\theta_3 (0) \theta_3 (r\pi /p)}{\theta_4 (0) \theta_4 (r\pi /p)}
\left[1-\frac{\theta_4' (r\pi /p)|\theta_2 (r\pi /p)|}
{\theta_3^2 (0) \theta_3 (r\pi /p)|\theta_1 (r\pi /p)|}\right]~.
\eeq
\beq\label{40}
\sum_{j=1}^{p} \frac{\theta_1 (z_j) \theta_1 (z_{j+r})}
{\theta_4 (z_j) \theta_4 (z_{j+r})} 
=\frac{p\theta_4' (r\pi /p) \theta_2 (r\pi /p)}
{\theta_2 (0) \theta_3 (0)
|\theta_1 (r\pi /p) \theta_2 (r\pi /p)|}~.
\eeq
\beq\label{41}
\sum_{j=1}^{p} \frac{\theta_2 (z_j) \theta_2 (z_{j+r})}
{\theta_4 (z_j) \theta_4 (z_{j+r})} 
=p\frac{\theta_2 (0) \theta_2 (r\pi /p)}{\theta_4 (0) \theta_4 (r\pi /p)}
\left[1-\frac{\theta_3 (r\pi /p)\theta_4' (r\pi /p)}
{\theta_2^2 (0) |\theta_1 (r\pi /p) \theta_2 (r\pi /p)|}\right]~.
\eeq
\beq\label{42}
\sum_{j=1}^{p} 
\frac{\theta_3 (z_j) \theta_3 (z_{j+1})...\theta_3 (z_{j+(l-1)})}
{\theta_4 (z_j) \theta_4 (z_{j+1})... \theta_4 (z_{j+l-1})} 
=\frac{p}{\pi} \int_{0}^{\pi} dz 
\frac{\theta_3 (z) \theta_3 (z+\pi /p)...\theta_3 (z+[l-1]\pi /p)}
{\theta_4 (z) \theta_4 (z+\pi /p)... \theta_4 (z+[l-1]\pi /p)}~,~(l \rm even)~.
\eeq
\beq\label{43}
\sum_{j=1}^{p} 
\frac{\theta_2 (z_j) \theta_3 (z_{j})}
{\theta_4^2 (z_j)}\left[\frac{\theta_1 (z_{j+r})}{\theta_4 (z_{j+r})}
+\frac{\theta_1 (z_{j-r})}{\theta_4 (z_{j-r})}\right] = 0~.
\eeq
\beq\label{44}
\sum_{j=1}^{p} 
\frac{\theta_1 (z_j) \theta_3 (z_{j})}
{\theta_4^2 (z_j)}\left[\frac{\theta_2 (z_{j+r})}{\theta_4 (z_{j+r})}
+\frac{\theta_2 (z_{j-r})}{\theta_4 (z_{j-r})}\right] = 0~.
\eeq
\beq\label{46}
\sum_{j=1}^{p} 
\frac{\theta_1 (z_j) \theta_2 (z_{j})}
{\theta_4^2 (z_j)}\left[\frac{\theta_3 (z_{j+r})}{\theta_4 (z_{j+r})}
+\frac{\theta_3 (z_{j-r})}{\theta_4 (z_{j-r})}\right] = 0~.
\eeq
\beq\label{45}
\sum_{j=1}^{p} 
\frac{\theta_2 (z_j)}
{\theta_4 (z_j)}\left[\frac{\theta_1 (z_{j+s}) \theta_3 (z_{j+r})}
{\theta_4 (z_{j+r}) \theta_4 (z_{j+s})}
+\frac{\theta_1 (z_{j-s}) \theta_3 (z_{j-r})}
{\theta_4 (z_{j-r}) \theta_4 (z_{j-s})}\right] =0~.
\eeq

We now write down results for those cyclic identities in IIa in which the
right hand side contained a definite integral which we could not then
evaluate. However, subsequently, we derived local identities 
\cite{klspr} using which we were able to evaluate all these integrals.
We are therefore giving answers using results in \cite{klspr}. 

\beqa\label{47}
&&\sum_{j=1}^{p} 
\frac{\theta_3^2 (z_j) \theta_3^2 (z_{j+r})}
{\theta_4^2 (z_j) \theta_4^2 (z_{j+r})}=
-2\frac{\theta_2^2 (r\pi /p)}{\theta_1^2 (r\pi /p)}
\sum_{j=1}^{p} \frac{\theta_3^2 (z_{j})}{\theta_4^2 (z_j)} \nonumber \\
&&+p\bigg [\frac{\theta_3^2 (0) \theta_2^2 (r\pi /p)}{\theta_4^2 (0) 
\theta_1^2 (r\pi /p)}+\frac{\theta_3^4 (0) \theta_3^2 (r\pi /p)}
{\theta_2^2 (0) \theta_4^2 (0) \theta_1^2 (r\pi /p)}-\frac{2\theta_2^2
(0) \theta_2 (r\pi /p) \theta_3 (r\pi /p)\theta_4'(r\pi /p)}
{\theta_3^2 (0) \theta_4^2 (0) \theta_1^3 (r\pi /p)} \bigg ]\,.
\eeqa
\beqa\label{48}
&&\sum_{j=1}^{p} 
\frac{\theta_1 (z_j) \theta_2 (z_j)}
{\theta_4^2 (z_j)}\left[\frac{\theta_1 (z_{j+r}) \theta_2 (z_{j+r})}
{\theta_4^2 (z_{j+r})}
+\frac{\theta_1 (z_{j-r}) \theta_2 (z_{j-r})}
{\theta_4^2 (z_{j-r})}\right] \nonumber \\
&&=4\frac{\theta_3 (0) \theta_4 (0) \theta_3 (r\pi /p) \theta_4 (r\pi /p)}
{\theta_2^2 (0) \theta_1^2 (r\pi /p)}
\sum_{j=1}^{p} \frac{\theta_3^2 (z_{j})}{\theta_4^2 (z_j)} \nonumber
\\
&&-2p\frac{\theta_3^4 (0) \theta_4 (r\pi /p)}{\theta_2^2 (0) \theta_4
(0) \theta_1^2 (r\pi /p)}\big [1+\frac{\theta_4^2 (0) \theta_3^2 (r\pi
/p)}{\theta_3^2 (0) \theta_4^2 (r\pi /p)} \big ]\big[\frac{\theta_3
(r\pi /p)}{\theta_3 (0)} - \frac{\theta_2 (r\pi /p) \theta_4'(r\pi
/p)}{\theta_3^3 (0) \theta_1 (r\pi /p)} \big ]\,.
\eeqa
\beqa\label{49}
&&\sum_{j=1}^{p} 
\frac{\theta_2 (z_j) \theta_3 (z_j)}
{\theta_4^2 (z_j)}\left[\frac{\theta_2 (z_{j+r}) \theta_3 (z_{j+r})}
{\theta_4^2 (z_{j+r})}
+\frac{\theta_2 (z_{j-r}) \theta_3 (z_{j-r})}
{\theta_4^2 (z_{j-r})}\right] \nonumber \\
&&=-4\frac{\theta_3 (0) \theta_2 (r\pi /p) \theta_3 (r\pi /p)}
{\theta_2 (0) \theta_1^2 (r\pi /p)}
\sum_{j=1}^{p} \frac{\theta_3^2 (z_{j})}{\theta_4^2 (z_j)} \nonumber
\\ 
&&+2p\frac{\theta_3^4 (0)}{\theta_4^2 (0) 
\theta_1^2 (r\pi /p)}\bigg [\frac{2 \theta_3 (r\pi
/p) \theta_2 (r\pi /p)}{\theta_2 (0) \theta_3 (0)} -\frac{\theta_4
' (r\pi /p) \theta_2 (0)}{\theta_1 (r\pi /p) \theta_3^3 (0)}
\big (\frac{\theta_3^2 (r\pi /p)}{\theta_3^2 (0)}
+ \frac{\theta_2^2 (r\pi /p)}{\theta_2^2 (0)} \big ) \bigg ]\,.
\eeqa
\beqa\label{50}
&&\sum_{j=1}^{p} 
\frac{\theta_1 (z_j) \theta_3 (z_j)}
{\theta_4^2 (z_j)}\left[\frac{\theta_1 (z_{j+r}) \theta_3 (z_{j+r})}
{\theta_4^2 (z_{j+r})}
+\frac{\theta_1 (z_{j-r}) \theta_3 (z_{j-r})}
{\theta_4^2 (z_{j-r})}\right] \nonumber \\ 
&&=4\frac{\theta_4 (0) \theta_2 (r\pi /p) \theta_4 (r\pi /p)}
{\theta_2 (0) \theta_1^2 (r\pi /p)}
\sum_{j=1}^{p} \frac{\theta_3^2 (z_{j})}{\theta_4^2 (z_j)} 
-2p\frac{\theta_4^3 (0) \theta_2 (0) \theta_2 (r\pi /p)}{\theta_3^2 (0) 
\theta_1^2 (r\pi /p) \theta_4 (r\pi /p)}\bigg [\big (\frac{\theta_3^2 (r\pi
/p)}{\theta_3^2 (0)} \nonumber \\
&&+\frac{\theta_4^2 (r\pi /p)}{\theta_4^2 (0)} \big
) -\frac{\theta_4
' (r\pi /p) \theta_2^2 (0) \theta_3 (r\pi /p)}{\theta_1 (r\pi /p)
\theta_3^4 (0) \theta_2 (r\pi /p)}
\big (\frac{\theta_2^2 (r\pi /p)}{\theta_2^2 (0)}
+ \frac{\theta_4^2 (r\pi /p)}{\theta_4^2 (0)} \big ) \bigg ]\,.
\eeqa
\beqa\label{53}
&&\sum_{j=1}^{p} 
\frac{\theta_3^3 (z_j)}
{\theta_4^3 (z_j)}\left[\frac{\theta_3 (z_{j+r})}
{\theta_4 (z_{j+r})}
+\frac{\theta_3 (z_{j-r})}
{\theta_4 (z_{j-r})}\right]  
=2\frac{\theta_2^2 (0) \theta_3 (r\pi /p) \theta_4 (r\pi /p)}
{\theta_3 (0) \theta_4 (0) \theta_1^2 (r\pi /p)}
\sum_{j=1}^{p} \frac{\theta_3^2 (z_{j})}{\theta_4^2 (z_j)} \nonumber \\ 
&&-2p\frac{\theta_3^2 (0)\theta_2^2 (r\pi /p)}{\theta_4 (0) \theta_4
(r\pi /p) \theta_1^2 (4\pi /p)} \big [\frac{\theta_3 (r\pi
/p)}{\theta_3 (0)}-\frac{\theta_2 (r\pi /p) \theta_4'(r\pi /p)}
{\theta_3^3 (0) \theta_1 (r\pi /p)} \big ]\,.
\eeqa
\beqa\label{54}
&&\sum_{j=1}^{p} 
\frac{\theta_1^3 (z_j)}
{\theta_4^3 (z_j)}\left[\frac{\theta_1 (z_{j+r})}
{\theta_4 (z_{j+r})}
+\frac{\theta_1 (z_{j-r})}
{\theta_4 (z_{j-r})}\right]  
=2\frac{\theta_4^4 (0) \theta_2 (r\pi /p) \theta_3 (r\pi /p)}
{\theta_2^3 (0) \theta_3 (0) \theta_1^2 (r\pi /p)}
\sum_{j=1}^{p} \frac{\theta_3^2 (z_{j})}{\theta_4^2 (z_j)} \nonumber \\ 
&&-2p\frac{\theta_3^2 (0)\theta_4^2 (0)}{\theta_2^2 (0) 
\theta_1^2 (r\pi /p)} \bigg [\frac{\theta_3 (r\pi
/p) \theta_2 (r\pi /p)}{\theta_3 (0) \theta_2 (0)} -\frac{\theta_4^2 (r\pi /p) 
\theta_4'(r\pi /p) \theta_2 (0)}
{\theta_4^2 (0) \theta_3^3 (0) \theta_1 (r\pi /p)} \bigg ]\,.
\eeqa
\beqa\label{55}
&&\sum_{j=1}^{p} 
\frac{\theta_2^3 (z_j)}
{\theta_4^3 (z_j)}\left[\frac{\theta_2 (z_{j+r})}
{\theta_4 (z_{j+r})}
+\frac{\theta_2 (z_{j-r})}
{\theta_4 (z_{j-r})}\right]  
=2\frac{\theta_3^4 (0) \theta_2 (r\pi /p) \theta_4 (r\pi /p)}
{\theta_2^3 (0) \theta_4 (0) \theta_1^2 (r\pi /p)}
\sum_{j=1}^{p} \frac{\theta_3^2 (z_{j})}{\theta_4^2 (z_j)} \nonumber \\ 
&&+2p\frac{\theta_2^3 (0)\theta_4 (0) \theta_2 (r\pi /p)}{\theta_3^4 (0) 
\theta_4 (r\pi /p)} \bigg [1-\frac{\theta_3^6 (0) \theta_4^2 (r\pi
/p)}{\theta_2^6 (0) \theta_1^2 (r\pi /p)} +\frac{\theta_3^3 (r\pi /p) 
\theta_4'(r\pi /p) \theta_4^2 (0)}
{\theta_2^4 (0) \theta_2 (r\pi /p) \theta_1^3 (r\pi /p)} \bigg ]\,.
\eeqa
\beq\label{59}
\sum_{j=1}^{p} 
\frac{\theta_1 (z_j) \theta_2 (z_j) \theta_3 (z_j)}
{\theta_4^3 (z_j)}\left[\frac{\theta_3^2 (z_{j+r})}
{\theta_4^2 (z_{j+r})}
+\frac{\theta_3^2 (z_{j-r})}
{\theta_4^2 (z_{j-r})}\right]  
=-2\frac{\theta_3^2 (0) \theta_2^2 (r\pi /p)}  
{\theta_4^2 (0) \theta_1^2 (r\pi /p)}
\sum_{j=1}^{p} \frac{\theta_1 (z_{j}) \theta_2 (z_j) \theta_3 (z_j)}
{\theta_4^3 (z_j)}~. 
\eeq
\beq\label{60}
\sum_{j=1}^{p} 
\frac{\theta_1^2 (z_j) \theta_2 (z_j) \theta_3 (z_j)}
{\theta_4^4 (z_j)}\left[\frac{\theta_1 (z_{j+r})}
{\theta_4 (z_{j+r})}
+\frac{\theta_1 (z_{j-r})}
{\theta_4 (z_{j-r})}\right]  
=-2\frac{\theta_4^2 (0) \theta_2 (r\pi /p) \theta_3 (r\pi /p)}  
{\theta_2 (0) \theta_3 (0) \theta_1^2 (r\pi /p)}
\sum_{j=1}^{p} \frac{\theta_1 (z_{j}) \theta_2 (z_j) \theta_3 (z_j)}
{\theta_4^3 (z_j)}~. 
\eeq
\beq\label{61}
\sum_{j=1}^{p} 
\frac{\theta_2^2 (z_j) \theta_1 (z_j) \theta_3 (z_j)}
{\theta_4^4 (z_j)}\left[\frac{\theta_2 (z_{j+r})}
{\theta_4 (z_{j+r})}
+\frac{\theta_2 (z_{j-r})}
{\theta_4 (z_{j-r})}\right]   
=2\frac{\theta_3^2 (0) \theta_2 (r\pi /p) \theta_4 (r\pi /p)}  
{\theta_2 (0) \theta_4 (0) \theta_1^2 (r\pi /p)}
\sum_{j=1}^{p} \frac{\theta_1 (z_{j}) \theta_2 (z_j) \theta_3 (z_j)}
{\theta_4^3 (z_j)}~. 
\eeq
\beqa\label{62}
&&\sum_{j=1}^{p} 
\frac{\theta_3^2 (z_j) \theta_1 (z_j) \theta_2 (z_j)}
{\theta_4^4 (z_j)}\left[\frac{\theta_3^3 (z_{j+r})}
{\theta_4^3 (z_{j+r})}
+\frac{\theta_3^3 (z_{j-r})}
{\theta_4^3 (z_{j-r})}\right] \nonumber \\  
&&=-4\frac{\theta_2^2 (0) \theta_2^2 (r\pi /p) \theta_3 (r\pi /p) 
\theta_4 (r\pi /p)}  
{\theta_3 (0) \theta_4 (0) \theta_1^4 (r\pi /p)}
\sum_{j=1}^{p} \frac{\theta_1 (z_{j}) \theta_2 (z_j) \theta_3 (z_j)}
{\theta_4^3 (z_j)}~. 
\eeqa
\beqa\label{63}
&&\sum_{j=1}^{p} 
\frac{\theta_1^2 (z_j) \theta_2 (z_j) \theta_3 (z_j)}
{\theta_4^4 (z_j)}\left[\frac{\theta_1^3 (z_{j+r})}
{\theta_4^3 (z_{j+r})}
+\frac{\theta_1^3 (z_{j-r})}
{\theta_4^3 (z_{j-r})}\right] \nonumber \\  
&&=-4\frac{\theta_4^2 (0) \theta_4^2 (r\pi /p) \theta_3 (r\pi /p) 
\theta_2 (r\pi /p)}  
{\theta_2 (0) \theta_3 (0) \theta_1^4 (r\pi /p)}
\sum_{j=1}^{p} \frac{\theta_1 (z_{j}) \theta_2 (z_j) \theta_3 (z_j)}
{\theta_4^3 (z_j)}~. 
\eeqa
\beqa\label{64}
&&\sum_{j=1}^{p} 
\frac{\theta_2^2 (z_j) \theta_1 (z_j) \theta_3 (z_j)}
{\theta_4^4 (z_j)}\left[\frac{\theta_2^3 (z_{j+r})}
{\theta_4^3 (z_{j+r})}
+\frac{\theta_2^3 (z_{j-r})}
{\theta_4^3 (z_{j-r})}\right] \nonumber \\  
&&=-4\frac{\theta_3^2 (0) \theta_3^2 (r\pi /p) \theta_2 (r\pi /p) 
\theta_4 (r\pi /p)}  
{\theta_2 (0) \theta_4 (0) \theta_1^4 (r\pi /p)}
\sum_{j=1}^{p} \frac{\theta_1 (z_{j}) \theta_2 (z_j) \theta_3 (z_j)}
{\theta_4^3 (z_j)}~. 
\eeqa
\beqa\label{65}
&&\sum_{j=1}^{p} 
\frac{\theta_1 (z_j) \theta_2 (z_j) \theta_3 (z_j)}
{\theta_4^3 (z_j)}\left[\frac{\theta_3^4 (z_{j+r})}
{\theta_4^4 (z_{j+r})}
+\frac{\theta_3^4 (z_{j-r})}
{\theta_4^4 (z_{j-r})}\right]  
=2\frac{\theta_2^2 (0) \theta_2^2 (r\pi /p)}{\theta_1^4 (4\pi /p)} \nonumber \\ 
&&\left [\frac{\theta_2^2 (r\pi /p)}{\theta_2^2 (0)}- 
\frac{\theta_2^2 (0) \theta_3^2 (r\pi /p) \theta_4^2 (r\pi /p)}
{\theta_3^2 (0) \theta_4^2 (0) \theta_2^2 (r\pi /p)}  
-\frac{\theta_3^2 (r\pi /p)}
{\theta_3^2 (0)}  
-\frac{\theta_4^2 (r\pi /p)}
{\theta_4^2 (0)}\right ]  
\sum_{j=1}^{p} \frac{\theta_1 (z_{j}) \theta_2 (z_j) \theta_3 (z_j)}
{\theta_4^3 (z_j)}~. 
\eeqa

\sss

\subsection{MI-II Identities with alternating signs.}

Let us now write the MI-II identities with alternating signs as given by 
Eqs. (219) to (239) [except identities (231), (232), (233) and (237)] of II in terms 
of Jacobi theta functions. It should be noted here that in this case $p$ is
necessarily even and that $r,s$ are therefore odd integers 
co-prime to $p$.

\beq\label{66}
\sum_{j=1}^{p} (-1)^{j-1} \frac{\theta_3 (z_j)
\theta_3 (z_{j+r})}
{\theta_4 (z_{j})
\theta_4 (z_{j+r})}  
=\frac{2}{\theta_3 (0) \theta_4 (0)}
\frac{\theta_2 (r\pi /p)}
{\theta_1 (r\pi /p)} 
\sum_{j=1}^{p} (-1)^{j-1} \frac{\theta_4' (z_j)}{\theta_4 (z_j)}~.
\eeq
\beq\label{67}
\sum_{j=1}^{p} (-1)^{j-1} \frac{\theta_1 (z_j)
\theta_1 (z_{j+r})}
{\theta_4 (z_{j})
\theta_4 (z_{j+r})}  
=\frac{2}{\theta_2 (0) \theta_3 (0)}
\frac{\theta_4 (r\pi /p)}
{\theta_1 (r\pi /p)} 
\sum_{j=1}^{p} (-1)^{j-1} \frac{\theta_4' (z_j)}{\theta_4 (z_j)}~.
\eeq
\beq\label{68}
\sum_{j=1}^{p} (-1)^{j-1} \frac{\theta_2 (z_j)
\theta_2 (z_{j+r})}
{\theta_4 (z_{j})
\theta_4 (z_{j+r})}  
=-\frac{2}{\theta_2 (0) \theta_4 (0)}
\frac{\theta_3 (r\pi /p)}
{\theta_1 (r\pi /p)} 
\sum_{j=1}^{p} (-1)^{j-1} \frac{\theta_4' (z_j)}{\theta_4 (z_j)}~.
\eeq
\beqa\label{69}
&&\sum_{j=1}^{p} (-1)^{j-1} \frac{\theta_3 (z_j) \theta_3 (z_{j+r}) 
\theta_3 (z_{j+2r}) \theta_3 (z_{j+3r})}
{\theta_4 (z_{j}) \theta_4 (z_{j+r}) \theta_4 (z_{j+2r}) \theta_4 (z_{j+3r})} \nonumber \\
&&=\frac{2}{\theta_3 (0) \theta_4 (0)} 
\left[\frac{\theta_2 (r\pi /p) \theta_2 (2r\pi /p) \theta_2 (3r\pi /p)}
{\theta_1 (r\pi /p) \theta_1 (2r\pi /p) \theta_1 (3r\pi /p)}+
\frac{\theta_2^2 (r\pi /p) \theta_2 (2r\pi /p)}
{\theta_1^2 (r\pi /p) \theta_1 (2r\pi /p)}\right]
\sum_{j=1}^{p} (-1)^{j-1} \frac{\theta_4' (z_j)}
{\theta_4 (z_j)}~.
\eeqa
This generalizes for any even number $l < p$ to:
\beqa\label{70}
&&\sum_{j=1}^{p} (-1)^{j-1} \frac{\theta_3 (z_j) \theta_3 (z_{j+r}) 
... \theta_3 (z_{j+(l-1)r})}
{\theta_4 (z_{j}) \theta_4 (z_{j+r})... \theta_4 (z_{j+(l-1)r})} \nonumber \\
&&=(-1)^{l/2} \frac{2}{\theta_3^2 (0)} 
\left(\sum_{k=1}^{l/2} (-1)^{k-1} \, \Pi_{n=1,n\ne k}^{l} 
\frac{\theta_2 ([n-k]r\pi /p)}
{\theta_1 ([n-k]r\pi /p)}\right)
\sum_{j=1}^{p} (-1)^{j-1} \frac{\theta_4' (z_j)}
{\theta_4 (z_j)}~.
\eeqa
Similarly for any even integer $l \le p$, $\theta_1$ and $\theta_2$ functions
satisfy the identities ($p \ge 4$)
\beqa\label{71}
&&\sum_{j=1}^{p} (-1)^{j-1} \frac{\theta_1 (z_j) \theta_1 (z_{j+r}) 
... \theta_1 (z_{j+[l-1]r})}
{\theta_4 (z_{j}) \theta_4 (z_{j+r})... \theta_4 (z_{j+[l-1]r})} \nonumber \\
&&=\frac{2}{\theta_3^2 (0)} 
\left(\sum_{k=1}^{l/2} (-1)^{k-1} \, \Pi_{n=1,n\ne k}^{l} 
\frac{\theta_4 ([n-k]r\pi /p)}
{\theta_1 ([n-k]r\pi /p)}\right)
\sum_{j=1}^{p} (-1)^{j-1} \frac{\theta_4' (z_j)}
{\theta_4 (z_j)}~.
\eeqa
\beqa\label{72}
&&\sum_{j=1}^{p} (-1)^{j-1} \frac{\theta_2 (z_j) \theta_2 (z_{j+r}) 
... \theta_2 (z_{j+[l-1]r})}
{\theta_4 (z_{j}) \theta_4 (z_{j+r})... \theta_4 (z_{j+[l-1]r})} \nonumber \\
&&=(-1)^{l/2} \frac{2}{\theta_3^2 (0)}
\left(\frac{\theta_3 (0)}{\theta_2 (0)}\right)^{2l} 
\left(\sum_{k=1}^{l/2} (-1)^{k-1} \, \Pi_{n=1,n\ne k}^{l} 
\frac{\theta_3 ([n-k]r\pi /p)}
{\theta_1 ([n-k]r\pi /p)}\right)
\sum_{j=1}^{p} (-1)^{j-1} \frac{\theta_4' (z_j)}
{\theta_4 (z_j)}~.
\eeqa
When $l=p$ $ (p \ge 4)$, the last two identities reduce to
\beq\label{73}
\Pi_{j=1}^{p} \frac{\theta_1 (z_j)} 
{\theta_4 (z_{j})} 
=\frac{1}{\theta_2^2 (0)} 
\left(\Pi_{n=1}^{\frac{(p-2)}{2}} 
\frac{\theta_4^2 (n\pi /p)}
{\theta_1^2 (n\pi /p)}\right)
\sum_{j=1}^{p} (-1)^{j-1} \frac{\theta_4' (z_j)}
{\theta_4 (z_j)}~.
\eeq
\beq\label{74}
\Pi_{j=1}^{p} \frac{\theta_2 (z_j)} 
{\theta_4 (z_{j})} 
=(-1)^{p/2} \frac{1}{\theta_2^2 (0)} 
\left(\Pi_{n=1}^{\frac{(p-2)}{2}} 
\frac{\theta_3^2 (n\pi /p)}
{\theta_1^2 (n\pi /p)}\right)
\sum_{j=1}^{p} (-1)^{j-1} \frac{\theta_4' (z_j)}
{\theta_4 (z_j)}~.
\eeq
\beq\label{75}
\sum_{j=1}^{p} (-1)^{j-1} \frac{\theta_3 (z_j)}
{\theta_4 (z_{j})}\left[\frac{\theta_1 (z_{j+r}) \theta_2 (z_{j+r})}
{\theta_4^2 (z_{j+r})} +
\frac{\theta_1 (z_{j-r}) \theta_2 (z_{j-r})}
{\theta_4^2 (z_{j-r})}\right]  
=-4\frac{\theta_3 (r\pi /p) \theta_4 (r\pi /p)}
{\theta_3 (0) \theta_4 (0) \theta_1^2 (r\pi /p)}
\sum_{j=1}^{p} (-1)^{j-1} \frac{\theta_4' (z_j)}{\theta_4 (z_j)}~.
\eeq
\beq\label{76}
\sum_{j=1}^{p} (-1)^{j-1} \frac{\theta_1 (z_j)}
{\theta_4 (z_{j})}\left[\frac{\theta_2 (z_{j+r}) \theta_3 (z_{j+r})}
{\theta_4^2 (z_{j+r})} +
\frac{\theta_2 (z_{j-r}) \theta_3 (z_{j-r})}
{\theta_4^2 (z_{j-r})}\right]  
=-4\frac{\theta_2 (r\pi /p) \theta_3 (r\pi /p)}
{\theta_2 (0) \theta_3 (0) \theta_1^2 (r\pi /p)}
\sum_{j=1}^{p} (-1)^{j-1} \frac{\theta_4' (z_j)}{\theta_4 (z_j)}~.
\eeq
\beq\label{77}
\sum_{j=1}^{p} (-1)^{j-1} \frac{\theta_2 (z_j)}
{\theta_4 (z_{j})}\left[\frac{\theta_1 (z_{j+r}) \theta_3 (z_{j+r})}
{\theta_4^2 (z_{j+r})} +
\frac{\theta_1 (z_{j-r}) \theta_3 (z_{j-r})}
{\theta_4^2 (z_{j-r})}\right]  
=-4\frac{\theta_2 (r\pi /p) \theta_4 (r\pi /p)}
{\theta_2 (0) \theta_4 (0) \theta_1^2 (r\pi /p)}
\sum_{j=1}^{p} (-1)^{j-1} \frac{\theta_4' (z_j)}{\theta_4 (z_j)}~.
\eeq
\beq\label{78}
\sum_{j=1}^{p} (-1)^{j-1} \frac{\theta_3^3 (z_j)}
{\theta_4^3 (z_{j})}\left[\frac{\theta_3 (z_{j+r})}
{\theta_4 (z_{j+r})} +
\frac{\theta_3 (z_{j-r})}
{\theta_4 (z_{j-r})}\right]  
=2\frac{\theta_2^2 (0) \theta_3 (0) \theta_3 (r\pi /p) \theta_4 (r\pi /p)}
{\theta_4^3 (0) \theta_1^2 (r\pi /p)}
\sum_{j=1}^{p} (-1)^{j-1} \frac{\theta_3^2 (z_j)}{\theta_4^2 (z_j)}~.
\eeq
\beq\label{78a}
\sum_{j=1}^{p} (-1)^{j-1} \frac{\theta_2^3 (z_j)}
{\theta_4^3 (z_{j})}\left[\frac{\theta_2 (z_{j+r})}
{\theta_4 (z_{j+r})} +
\frac{\theta_2 (z_{j-r})}
{\theta_4 (z_{j-r})}\right]  
=2\frac{\theta_3^4 (0) \theta_2 (r\pi /p) \theta_4 (r\pi /p)}
{\theta_2^3 (0) \theta_4 (0) \theta_1^2 (r\pi /p)}
\sum_{j=1}^{p} (-1)^{j-1} \frac{\theta_3^2 (z_j)}{\theta_4^2 (z_j)}~.
\eeq
\beq\label{79}
\sum_{j=1}^{p} (-1)^{j-1} \frac{\theta_1^3 (z_j)}
{\theta_4^3 (z_{j})}\left[\frac{\theta_1 (z_{j+r})}
{\theta_4 (z_{j+r})} +
\frac{\theta_1 (z_{j-r})}
{\theta_4 (z_{j-r})}\right]  
=2\frac{\theta_4^4 (0) \theta_2 (r\pi /p) \theta_3 (r\pi /p)}
{\theta_2^3 (0) \theta_3 (0) \theta_1^2 (r\pi /p)}
\sum_{j=1}^{p} (-1)^{j-1} \frac{\theta_3^2 (z_j)}{\theta_4^2 (z_j)}~.
\eeq
\beqa\label{80}
&&\sum_{j=1}^{p} (-1)^{j-1} \frac{\theta_3^3 (z_j)}
{\theta_4^3 (z_{j})}\left[\frac{\theta_1 (z_{j+r}) \theta_2 (z_{j+r})}
{\theta_4^2 (z_{j+r})} +
\frac{\theta_1 (z_{j-r}) \theta_2 (z_{j-r})}
{\theta_4^2 (z_{j-r})}\right] \nonumber \\  
&&=-2\frac{\theta_2^2 (0) \theta_3 (r\pi /p) \theta_4 (r\pi /p)}
{\theta_3 (0) \theta_4 (0) \theta_1^2 (r\pi /p)}
\sum_{j=1}^{p} (-1)^{j-1} \frac{\theta_1 (z_j) \theta_2 (z_j) \theta_3 (z_j)}
{\theta_4^3 (z_j)} \nonumber \\
&&-12 \frac{\theta_2^2 (r\pi /p) \theta_3 (r\pi /p) 
\theta_4 (r\pi /p)}{\theta_3 (0) \theta_4 (0) \theta_1^4 (r\pi /p)}
\sum_{j=1}^{p} (-1)^{j-1} \frac{\theta_4' (z_j)}{\theta_4 (z_j)}~.
\eeqa
\beqa\label{81}
&&\sum_{j=1}^{p} (-1)^{j-1} 
\frac{\theta_2^2 (z_j) \theta_1 (z_j) \theta_3 (z_j)}
{\theta_4^4 (z_{j})}\left[\frac{\theta_2 (z_{j+r})}
{\theta_4 (z_{j+r})} +
\frac{\theta_2 (z_{j-r})}
{\theta_4 (z_{j-r})}\right] \nonumber \\  
&&=2\frac{\theta_3^2 (0) \theta_2 (r\pi /p) \theta_4 (r\pi /p)}
{\theta_2 (0) \theta_4 (0) \theta_1^2 (r\pi /p)}
\sum_{j=1}^{p} (-1)^{j-1} \frac{\theta_1 (z_j) \theta_2 (z_j) \theta_3 (z_j)}
{\theta_4^3 (z_j)} \nonumber \\
&&-4 \frac{\theta_3^2 (r\pi /p) \theta_2 (r\pi /p) 
\theta_4 (r\pi /p)}{\theta_2 (0) \theta_4 (0) \theta_1^4 (r\pi /p)}
\sum_{j=1}^{p} (-1)^{j-1} \frac{\theta_4' (z_j)}{\theta_4 (z_j)}~.
\eeqa

\section{Identities following from MI-III.}

We shall now rewrite the identities following from master identity MI-III in
terms of theta functions. It
is worth recalling that unlike the previous two sections, the period for 
these identities,
as well as those in the next section 
is $2\pi$. Further, $p$ is necessarily an odd integer. 
The identities (144) to (170) [except identities
(152),(154),(161),(163),(164) and (168)] of IIa, when
re-expressed in terms of Jacobi theta functions are given below.

\beq\label{83}
\sum_{j=1}^{p} \frac{\theta_2 (z_j)}{\theta_4 (z_j)}\left[
\frac{\theta_3 (z_{j+r})}{\theta_4 (z_{j+r})}+
\frac{\theta_3 (z_{j-r})}{\theta_4 (z_{j-r})}\right] = 0~.
\eeq
\beqa\label{84}
&&\sum_{j=1}^{p} \frac{\theta_1 (z_j)}{\theta_4 (z_j)}\frac{\theta_1 (z_{j+r})}
{\theta_4 (z_{j+r})}...\frac{\theta_1 (z_{j+(l-1)r})}{\theta_4 (z_{j+(l-1)r})} 
=\bigg [(-1)^{(l-1)/2} \Pi_{k=1}^{\frac{l-1}{2}} \frac{\theta_4^2 (2rk\pi /p)}
{\theta_1^2 (2rk\pi /p)} \nonumber \\
&&+2\sum_{k=1}^{(l-1)/2} \Pi_{n\ne k,n =1}^{l} \frac{\theta_4 (2[n-k]r\pi /p)}
{\theta_1 (2[n-k]r \pi /p)} \bigg ] 
\sum_{j=1}^{p} \frac{\theta_1 (z_j)}{\theta_4 (z_j)}~,
\eeqa
where $l$ is any odd integer ($3 \le l \le p$).
Note that unlike the last two sections, here $z_j \equiv
z+\frac{2(j-1)\pi}{K}$ where as before $z \equiv
\frac{u\pi}{2K}=\frac{u}{\theta^2_3 (0)}$.
In the special case when $l=p$, this identity takes the 
elegant form
\beq\label{85}
\Pi_{j=1}^{p} \frac{\theta_1 (z_j)}{\theta_4 (z_j)} = 
(-1)^{\frac{(p-1)}{2}} \Pi_{n=1}^{\frac{p-1}{2}}
\frac{\theta_4^2 (2n\pi /p)}{\theta_1^2 (2n\pi /p)} \sum_{j=1}^{p} 
\frac{\theta_1 (z_j)}{\theta_4 (z_j)}~.
\eeq
\beqa\label{86}
&&\sum_{j=1}^{p} \frac{\theta_1^2 (z_j)}{\theta_4^2 (z_{j})}
\left[\frac{\theta_1 (z_{j+r})}{\theta_4 (z_{j+r})} +
\frac{\theta_1 (z_{j-r})}{\theta_4 (z_{j-r})}\right] \nonumber \\ 
&&= -2\frac{\theta_4^2 (0)}{\theta_1^2 (2r\pi /p)} 
\left[\frac{\theta_2 (2r\pi /p) \theta_3 (2r\pi /p)}
{\theta_2 (0) \theta_3 (0)} 
-\frac{\theta_4^2 (2r\pi /p)}{\theta_4^2 (0)}\right] 
\sum_{j=1}^{p} 
\frac{\theta_1 (z_j)}{\theta_4 (z_j)}~.
\eeqa
\beqa\label{87}
&&\sum_{j=1}^{p} \frac{\theta_2 (z_j)}{\theta_4 (z_{j})}
\left[\frac{\theta_1 (z_{j+r}) \theta_2 (z_{j+r})}{\theta_4^2 (z_{j+r})} +
\frac{\theta_1 (z_{j-r}) \theta_2 (z_{j-r})}{\theta_4^2 (z_{j-r})}\right] \nonumber \\ 
&&= 2\frac{\theta_3^2 (0)
\theta_4 (2r\pi /p)}{\theta_4 (0) \theta_1^2 (2r\pi /p)} 
\left[\frac{\theta_2 (2r\pi /p)}{\theta_2 (0)}
-\frac{\theta_3 (2r\pi /p)}{\theta_3 (0)}\right] 
\sum_{j=1}^{p} 
\frac{\theta_1 (z_j)}{\theta_4 (z_j)}~.
\eeqa
\beqa\label{88}
&&\sum_{j=1}^{p} \frac{\theta_3 (z_j)}{\theta_4 (z_{j})}
\left[\frac{\theta_1 (z_{j+r}) \theta_3 (z_{j+r})}{\theta_4^2 (z_{j+r})} +
\frac{\theta_1 (z_{j-r}) \theta_3 (z_{j-r})}{\theta_4^2 (z_{j-r})}\right] \nonumber \\ 
&&= 2\frac{\theta_2^2 (0)
\theta_4 (2r\pi /p)}{\theta_4 (0) \theta_1^2 (2r\pi /p)} 
\left[\frac{\theta_3 (2r\pi /p)}{\theta_3 (0)}
-\frac{\theta_2 (2r\pi /p)}{\theta_2 (0)}\right] 
\sum_{j=1}^{p} 
\frac{\theta_1 (z_j)}{\theta_4 (z_j)}~.
\eeqa
\beqa\label{89}
&&\sum_{j=1}^{p} \frac{\theta_1 (z_j)}{\theta_4 (z_{j})}
\left[\frac{\theta_2 (z_{j+r}) \theta_2 (z_{j+s})}
{\theta_4 (z_{j+r}) \theta_4 (z_{j+s})} +
\frac{\theta_2 (z_{j-r}) \theta_2 (z_{j-s})}
{\theta_4 (z_{j-r}) \theta_4 (z_{j-s})}\right]  
= -2
\bigg[\frac{\theta_3 (2r\pi /p) \theta_3 (2s\pi /p)}
{\theta_1 (2r\pi /p) \theta_1 (2s\pi /p)} \nonumber \\
&&+\frac{\theta_3 (0)}{\theta_4 (0)} 
\frac{\theta_3 (2[r-s]\pi /p)}{\theta_1 (2[r-s]\pi /p)}
\left(\frac{\theta_4 (2r\pi /p)}{\theta_1 (2r\pi /p)}
-\frac{\theta_4 (2s\pi /p)}{\theta_1 (2s\pi /p)}\right)
\bigg] 
\sum_{j=1}^{p} 
\frac{\theta_1 (z_j)}{\theta_4 (z_j)}~.
\eeqa
\beqa\label{90}
&&\sum_{j=1}^{p} \frac{\theta_1 (z_j)}{\theta_4 (z_{j})}
\left[\frac{\theta_3 (z_{j+r}) \theta_3 (z_{j+s})}
{\theta_4 (z_{j+r}) \theta_4 (z_{j+s})} +
\frac{\theta_3 (z_{j-r}) \theta_3 (z_{j-s})}
{\theta_4 (z_{j-r}) \theta_4 (z_{j-s})}\right]  
= -2
\bigg[\frac{\theta_2 (2r\pi /p) \theta_2 (2s\pi /p)}
{\theta_1 (2r\pi /p) \theta_1 (2s\pi /p)} \nonumber \\
&&+\frac{\theta_2 (0)}{\theta_4 (0)}\frac{\theta_2 (2[r-s]\pi /p)}
{\theta_1 (2[r-s]\pi /p)}
\left(\frac{\theta_4 (2r\pi /p)}{\theta_1 (2r\pi /p)}
-\frac{\theta_4 (2s\pi /p)}{\theta_1 (2s\pi /p)}\right)
\bigg] 
\sum_{j=1}^{p} 
\frac{\theta_1 (z_j)}{\theta_4 (z_j)}~.
\eeqa
\beqa\label{91}
&&\sum_{j=1}^{p} \frac{\theta_3 (z_j)}{\theta_4 (z_{j})}
\left[\frac{\theta_3 (z_{j+r}) \theta_1 (z_{j+s})}
{\theta_4 (z_{j+r}) \theta_4 (z_{j+s})} +
\frac{\theta_3 (z_{j-r}) \theta_1 (z_{j-s})}
{\theta_4 (z_{j-r}) \theta_4 (z_{j-s})}\right]  
= -2\frac{\theta_2 (2r\pi /p)}{\theta_1 (2r\pi /p)} 
\bigg[\frac{\theta_2 (2[r-s]\pi /p)}
{\theta_1 (2[r-s]\pi /p)} \nonumber \\
&&+\frac{\theta_2 (2s\pi /p) \theta_2 (0)}
{\theta_1 (2s\pi /p) \theta_3 (0)}
\left(\frac{\theta_4 (2r\pi /p)}{\theta_1 (2r\pi /p)}
-\frac{\theta_4 (2[r-s]\pi /p)}{\theta_1 (2[r-s]\pi /p)}\right)
\bigg] 
\sum_{j=1}^{p} 
\frac{\theta_1 (z_j)}{\theta_4 (z_j)}~.
\eeqa
\beqa\label{92}
&&\sum_{j=1}^{p} \frac{\theta_2 (z_j) \theta_3 (z_j)}{\theta_4^2 (z_{j})}
\left[\frac{\theta_1^2 (z_{j+r})}
{\theta_4^2 (z_{j+r})} +
\frac{\theta_1^2 (z_{j-r})}
{\theta_4^2 (z_{j-r})}\right] \nonumber \\ 
&&= 2 \frac{\theta_4^2 (2r\pi /p)}{\theta_1^2 (2r\pi /p)}
\left[1
+\frac{\theta_4^2 (0) \theta_2 (2r\pi /p) \theta_3 (2r\pi /p)}
{\theta_2 (0) \theta_3 (0) \theta_4^2 (2r\pi /p)}\right]
\sum_{j=1}^{p} \frac{\theta_2 (z_j) \theta_3 (z_j)}{\theta_4^2 (z_j)}~.
\eeqa
\beqa\label{93}
&&\sum_{j=1}^{p} \frac{\theta_1 (z_j) \theta_3 (z_j)}{\theta_4^2 (z_{j})}
\left[\frac{\theta_1 (z_{j+r}) \theta_2 (z_{j+r})}
{\theta_4^2 (z_{j+r})} +
\frac{\theta_1 (z_{j-r}) \theta_2 (z_{j-r})}
{\theta_4^2 (z_{j-r})}\right] \nonumber \\ 
&&= 2 \frac{\theta_4 (0)}{\theta_1 (2r\pi /p)}
\left[\frac{\theta_2 (2r\pi /p)}{\theta_2 (0)}
+\frac{\theta_3 (2r\pi /p)}
{\theta_3 (0)}\right]
\sum_{j=1}^{p} \frac{\theta_2 (z_j) \theta_3 (z_j)}{\theta_4^2 (z_j)}~.
\eeqa
\beqa\label{94}
&&\sum_{j=1}^{p} \frac{\theta_2 (z_j) \theta_3 (z_j)}{\theta_4^2 (z_{j})}
\left[\frac{\theta_1 (z_{j+r}) \theta_1 (z_{j+s})}
{\theta_4 (z_{j+r}) \theta_4 (z_{j+s})} +
\frac{\theta_1 (z_{j-r}) \theta_1 (z_{j-s})}
{\theta_4 (z_{j-r}) \theta_4 (z_{j-s})}\right] \nonumber \\
&&= 2 \frac{\theta_4 (2r\pi /p) \theta_4 (2s\pi /p)}
{\theta_1 (2r\pi /p) \theta_1 (2s\pi /p)}
\sum_{j=1}^{p} \frac{\theta_2 (z_j) \theta_3 (z_j)}{\theta_4^2 (z_j)}~.
\eeqa
\beqa\label{95}
&&\sum_{j=1}^{p} \frac{\theta_2 (z_j) \theta_3 (z_j)}{\theta_4^2 (z_{j})}
\left[\frac{\theta_2 (z_{j+r}) \theta_2 (z_{j+s})}
{\theta_4 (z_{j+r}) \theta_4 (z_{j+s})} +
\frac{\theta_2 (z_{j-r}) \theta_2 (z_{j-s})}
{\theta_4 (z_{j-r}) \theta_4 (z_{j-s})}\right] \nonumber \\
&&= -2 \frac{\theta_2^2 (0) \theta_3 (2r\pi /p) \theta_3 (2s\pi /p)}
{\theta_3^2 (0) \theta_1 (2r\pi /p) \theta_1 (2s\pi /p)}
\sum_{j=1}^{p} \frac{\theta_2 (z_j) \theta_3 (z_j)}{\theta_4^2 (z_j)}~.
\eeqa
\beqa\label{96}
&&\sum_{j=1}^{p} \frac{\theta_2 (z_j) \theta_3 (z_j)}{\theta_4^2 (z_{j})}
\left[\frac{\theta_3 (z_{j+r}) \theta_3 (z_{j+s})}
{\theta_4 (z_{j+r}) \theta_4 (z_{j+s})} +
\frac{\theta_3 (z_{j-r}) \theta_3 (z_{j-s})}
{\theta_4 (z_{j-r}) \theta_4 (z_{j-s})}\right] \nonumber \\
&&= -2 \frac{\theta_2 (2r\pi /p) \theta_2 (2s\pi /p)}
{\theta_1 (2r\pi /p) \theta_1 (2s\pi /p)}
\sum_{j=1}^{p} \frac{\theta_2 (z_j) \theta_3 (z_j)}{\theta_4^2 (z_j)}~.
\eeqa
\beqa\label{97}
&&\sum_{j=1}^{p} \frac{\theta_1 (z_j) \theta_2 (z_j)}{\theta_4^2 (z_{j})}
\left[\frac{\theta_3 (z_{j+r}) \theta_1 (z_{j+s})}
{\theta_4 (z_{j+r}) \theta_4 (z_{j+s})} +
\frac{\theta_3 (z_{j-r}) \theta_1 (z_{j-s})}
{\theta_4 (z_{j-r}) \theta_4 (z_{j-s})}\right] \nonumber \\
&&= 2 \frac{\theta_4 (0) \theta_2 (2r\pi /p) \theta_4 (2s\pi /p)}
{\theta_2 (0) \theta_1 (2r\pi /p) \theta_1 (2s\pi /p)}
\sum_{j=1}^{p} \frac{\theta_2 (z_j) \theta_3 (z_j)}{\theta_4^2 (z_j)}~.
\eeqa
\beqa\label{82}
&&\sum_{j=1}^{p} \frac{\theta_1 (z_j) \theta_2 (z_{j}) \theta_3 (z_{j})}
{\theta_4^3 (z_{j})} \left[\frac{\theta_1 (z_{j+r})}{\theta_4 (z_{j+r})} +
\frac{\theta_1 (z_{j-r})}{\theta_4 (z_{j-r})}\right] \nonumber \\ 
&&= -2\frac{\theta_4^2 (0)}
{\theta_2 (0) \theta_3 (0)} \frac{\theta_2 (2r\pi /p) \theta_3 (2r\pi /p)}
{\theta_1^2 (2r\pi /p)} \sum_{j=1}^{p} 
\frac{\theta_2 (z_j) \theta_3 (z_{j})}{\theta_4^2 (z_j)}~.
\eeqa
\beqa\label{98}
&&\sum_{j=1}^{p} \frac{\theta_1 (z_j) \theta_2 (z_j) \theta_3 (z_j)}
{\theta_4^3 (z_{j})}
\left[\frac{\theta_1^3 (z_{j+r})}
{\theta_4^3 (z_{j+r})} +
\frac{\theta_1^3 (z_{j-r})}
{\theta_4^3 (z_{j-r})}\right] \nonumber \\ 
&&= - \frac{2\theta_4^2 (0) \theta_4^2 (2r\pi /p)}{\theta_1^4(2r\pi /p)}
\bigg[\frac{\theta_2^2 (2r\pi /p)}{\theta_2^2 (0)}+
\frac{\theta_4^2 (0) \theta_2^2 (2r\pi /p) \theta_3^2 (2r\pi /p)}
{\theta_4^2 (2r\pi /p) \theta_2^2 (0) \theta_3^2 (0)}+
\frac{\theta_3^2 (2r\pi /p)}{\theta_3^2 (0)} \nonumber \\
&&+3\frac{\theta_2 (2r\pi /p) \theta_3 (2r\pi /p)}
{\theta_2 (0) \theta_3 (0)}\bigg] 
\sum_{j=1}^{p} \frac{\theta_2 (z_j) \theta_3 (z_j)}{\theta_4^2 (z_j)}~.
\eeqa
\beqa\label{99}
&&\sum_{j=1}^{p} \frac{\theta_1^4 (z_j)}
{\theta_4^4 (z_{j})}\left[\frac{\theta_1 (z_{j+r})}
{\theta_4 (z_{j+r})} +
\frac{\theta_1 (z_{j-r})}
{\theta_4 (z_{j-r})}\right]  
= -2 \frac{\theta_4^2 (0) \theta_2 (2r\pi /p) \theta_3 (2r\pi /p)}
{\theta_2 (0) \theta_3 (0) \theta_1^2 (2r\pi /p)}
\sum_{j=1}^{p} \frac{\theta_1^3 (z_j)}{\theta_4^3 (z_j)} \nonumber \\
&&+2\frac{\theta_2^2 (2r\pi /p) \theta_4^2 (0)}{\theta_1^4 (2r\pi /p)}
\left[\frac{\theta_4^2 (2r\pi /p)}{\theta_4^2 (0)}-
\frac{\theta_2 (2r\pi /p) \theta_3 (2r\pi /p)}
{\theta_2 (0) \theta_3 (0)}\right]
\sum_{j=1}^{p} \frac{\theta_1 (z_j)}{\theta_4 (z_j)}~.
\eeqa
\beqa\label{100}
&&\sum_{j=1}^{p} \frac{\theta_1^3 (z_j)}
{\theta_4^3 (z_{j})}\left[\frac{\theta_1^2 (z_{j+r})}
{\theta_4^2 (z_{j+r})} +
\frac{\theta_1^2 (z_{j-r})}
{\theta_4^2 (z_{j-r})}\right]  
= 2 \frac{\theta_4^2 (2r\pi /p)}
{\theta_1^2 (2r\pi /p)}
\sum_{j=1}^{p} \frac{\theta_1^3 (z_j)}{\theta_4^3 (z_j)} \nonumber \\
&&+2\frac{\theta_4^2 (0) \theta_4^2 (2r\pi /p)}{\theta_1^4 (2r\pi /p)}
\bigg[
\frac{\theta_2^2 (2r\pi /p)}{\theta_2^2 (0)}+
\frac{\theta_3^2 (2r\pi /p)}{\theta_3^2 (0)} \nonumber \\
&&+\frac{\theta_4^2 (0) \theta_2^2 (2r\pi /p) \theta_3^2 (2r\pi /p)}
{\theta_4^2 (2r\pi /p) \theta_2^2 (0) \theta_3^2 (0)}
-3\frac{\theta_2 (2r\pi /p) \theta_3 (2r\pi /p)}
{\theta_2 (0) \theta_3 (0)}\bigg]
\sum_{j=1}^{p} \frac{\theta_1 (z_j)}{\theta_4 (z_j)}~.
\eeqa
\beqa\label{100a}
&&\sum_{j=1}^{p} \frac{\theta_1^4 (z_j)}
{\theta_4^4 (z_{j})}\left[\frac{\theta_2 (z_{j+r}) \theta_3 (z_{j+r})}
{\theta_4^2 (z_{j+r})} +
\frac{\theta_2 (z_{j-r}) \theta_3 (z_{j-r})}
{\theta_4^2 (z_{j-r})}\right]  
= 2 \frac{\theta_4^2 (0) \theta_2 (2r\pi /p) \theta_3 (2r\pi /p)}
{\theta_2 (0) \theta_3 (0) \theta_1^2 (2r\pi /p)}
\sum_{j=1}^{p} \frac{\theta_1^2 (z_j) \theta_2 (z_{j}) \theta_3 (z_j)}
{\theta_4^4 (z_j)} \nonumber \\
&&+2\frac{\theta_4^2 (0) \theta_4^2 (2r\pi /p)}{\theta_1^4 (2r\pi /p)}
\left[
\frac{\theta_4^2 (2r\pi /p)}{\theta_4^2 (0)}+
\frac{3\theta_2 (2r\pi /p) \theta_3 (2r\pi /p)}{\theta_2 (0) \theta_3 (0)} 
\right]
\sum_{j=1}^{p} \frac{\theta_2 (z_j) \theta_3 (z_j)}{\theta_4^2 (z_j)}~.
\eeqa
\beqa\label{100b}
&&\sum_{j=1}^{p} \frac{\theta_1^4 (z_j)}
{\theta_4^4 (z_{j})}\left[\frac{\theta_2 (z_{j+s}) \theta_3 (z_{j+r})}
{\theta_4 (z_{j+r}) \theta_4 (z_{j+s})} +
\frac{\theta_2 (z_{j-s}) \theta_3 (z_{j-r})}
{\theta_4 (z_{j-r}) \theta_4 (z_{j-s})}\right]   
= 2 \frac{\theta_4^2 (0) \theta_2 (2r\pi /p) \theta_3 (2s\pi /p)}
{\theta_2 (0) \theta_3 (0) \theta_1 (2r\pi /p) \theta_1 (2s\pi /p)}
\sum_{j=1}^{p} \nonumber \\
&&\frac{\theta_1^2 (z_j) \theta_2 (z_{j}) \theta_3 (z_j)}
{\theta_4^4 (z_j)}  
+2\frac{\theta_4^2 (0)}{\theta_2 (0) \theta_3 (0)}
\bigg[ 
\frac{\theta_3 (2r\pi /p) \theta_4 (2r\pi /p) \theta_2 (2s\pi /p) 
\theta_4 (2s\pi /p)}{\theta_1^2 (2r\pi /p) \theta_1^2 (2s\pi /p)} \nonumber \\
&&+\frac{\theta_2 (2r\pi /p) \theta_3 (2s\pi /p)}
{\theta_1 (2r\pi /p) \theta_3 (2s\pi /p)}\left(\frac{\theta_4^2(2r\pi /p)}
{\theta_1^2 (2r\pi /p)}+\frac{\theta_4^2 (2s\pi /p)}
{\theta_1^2 (2s\pi /p)}\right) 
\bigg]
\sum_{j=1}^{p} \frac{\theta_2 (z_j) \theta_3 (z_j)}{\theta_4^2 (z_j)}~.
\eeqa

\sss

\section{Identities following from MI-IV.}

We shall now rewrite the identities following from master identity MI-IV in
terms of theta functions. As in the last section, the period for 
these identities is $2\pi$. Further, $p$ is necessarily an odd integer. 
The identities (171) to (194) [except identities (181), (183),
(186) and (192)] of IIa, when
re-expressed in terms of Jacobi theta functions are given below.
As in the previous section, in this section too, 
$z_j \equiv z+2(j-1)\pi /p$ with $z = u \pi /2K = u /\theta_3^2 (0)$.

\beq\label{102}
\sum_{j=1}^{p} \frac{\theta_1 (z_j) \theta_2 (z_{j}) \theta_3 (z_{j})}
{\theta_4^3 (z_{j})} \left[\frac{\theta_2 (z_{j+r})}{\theta_4 (z_{j+r})} +
\frac{\theta_2 (z_{j-r})}{\theta_4 (z_{j-r})}\right]  
= 2\frac{\theta_4 (0)}
{\theta_3 (0)} \frac{\theta_2 (2\pi /p) \theta_3 (2\pi /p)}
{\theta_1^2 (2\pi /p)} \sum_{j=1}^{p} 
\frac{\theta_1 (z_j) \theta_3 (z_{j})}{\theta_4^2 (z_j)}~.
\eeq
\beq\label{103}
\sum_{j=1}^{p} \frac{\theta_3 (z_j)}{\theta_4 (z_j)}\left[
\frac{\theta_1 (z_{j+r})}{\theta_4 (z_{j+r})}+
\frac{\theta_1 (z_{j-r})}{\theta_4 (z_{j-r})}\right] = 0~.
\eeq
\beqa\label{104}
&&\sum_{j=1}^{p} \frac{\theta_2 (z_j)}{\theta_4 (z_j)}\frac{\theta_2 (z_{j+r})}
{\theta_4 (z_{j+r})}...\frac{\theta_2 (z_{j+(l-1)r})}{\theta_4 (z_{j+(l-1)r})} 
=\bigg [\Pi_{k=1}^{\frac{(l-1)}{2}} \frac{\theta_3^2 (2rk\pi /p)}
{\theta_1^2 (2rk\pi /p)} \nonumber \\
&&+2(-1)^{\frac{(l-1)}{2}} \sum_{k=1}^{\frac{(l-1)}{2}} \Pi_{n\ne k,n =1}^{l} 
\frac{\theta_3 (2[n-k]r\pi /p)}
{\theta_1 (2[n-k]r \pi /p)} \bigg ] 
\sum_{j=1}^{p} \frac{\theta_2 (z_j)}{\theta_4 (z_j)}~,
\eeqa
where $l$ is any odd integer ($3 \le l \le p$).
In the special case when $l=p$, this identity takes the form
\beq\label{105}
\Pi_{j=1}^{p} \frac{\theta_2 (z_j)}{\theta_4 (z_j)} = 
\Pi_{n=1}^{\frac{(p-1)}{2}}
\frac{\theta_3^2 (2n\pi /p)}{\theta_1^2 (2n\pi /p)} \sum_{j=1}^{p} 
\frac{\theta_2 (z_j)}{\theta_4 (z_j)}~.
\eeq
\beqa\label{101}
&&\sum_{j=1}^{p} \frac{\theta_2^2 (z_{j})}
{\theta_4^2 (z_{j})} \left[\frac{\theta_2 (z_{j+r})}{\theta_4 (z_{j+r})} +
\frac{\theta_2 (z_{j-r})}{\theta_4 (z_{j-r})}\right] \nonumber \\ 
&&= 2\frac{\theta_4 (0)
\theta_2 (2r\pi /p) \theta_4 (2r\pi /p)}
{\theta_1^2 (2r\pi /p)} \left[\frac{\theta_2 (2r\pi /p)}{\theta_2 (0)}
-\frac{\theta_4 (0)
\theta_3^2 (2r\pi /p)}{\theta_3^2 (0) \theta_4 (2r\pi /p)}
\right]\sum_{j=1}^{p} 
\frac{\theta_2 (z_j)}{\theta_4 (z_j)}~.
\eeqa
\beqa\label{106}
&&\sum_{j=1}^{p} \frac{\theta_3 (z_j)}{\theta_4 (z_{j})}
\left[\frac{\theta_2 (z_{j+r}) \theta_3 (z_{j+r})}{\theta_4^2 (z_{j+r})} +
\frac{\theta_2 (z_{j-r}) \theta_3 (z_{j-r})}{\theta_4^2 (z_{j-r})}\right] \nonumber \\ 
&&= -2\frac{\theta_2^2 (0)
\theta_3 (2r\pi /p)}
{\theta_3 (0) \theta_1^2 (2r\pi /p)} 
\left[\frac{\theta_2 (2r\pi /p)}{\theta_2 (0)}
-\frac{\theta_4 (2r\pi /p)}{\theta_4 (0)}\right] 
\sum_{j=1}^{p} 
\frac{\theta_2 (z_j)}{\theta_4 (z_j)}~.
\eeqa
\beqa\label{107}
&&\sum_{j=1}^{p} \frac{\theta_1 (z_j)}{\theta_4 (z_{j})}
\left[\frac{\theta_1 (z_{j+r}) \theta_2 (z_{j+r})}{\theta_4^2 (z_{j+r})} +
\frac{\theta_1 (z_{j-r}) \theta_2 (z_{j-r})}{\theta_4^2 (z_{j-r})}\right] \nonumber \\ 
&&= -2\frac{\theta_4^2 (0)
\theta_3 (2r\pi /p)}
{\theta_3 (0) \theta_1^2 (2r\pi /p)} 
\left[\frac{\theta_2 (2r\pi /p)}{\theta_2 (0)}
-\frac{\theta_4 (2r\pi /p)}{\theta_4 (0)}\right] 
\sum_{j=1}^{p} 
\frac{\theta_2 (z_j)}{\theta_4 (z_j)}~.
\eeqa
\beqa\label{108}
&&\sum_{j=1}^{p} \frac{\theta_3 (z_j)}{\theta_4 (z_{j})}
\left[\frac{\theta_2 (z_{j+s}) \theta_3 (z_{j+r})}
{\theta_4 (z_{j+r}) \theta_4 (z_{j+s})} +
\frac{\theta_2 (z_{j-s}) \theta_3 (z_{j-r})}
{\theta_4 (z_{j-r}) \theta_4 (z_{j-s})}\right]  
= -2
\bigg[\frac{\theta_2 (2[r-s]\pi /p) \theta_2 (2r\pi /p)}
{\theta_1 (2r\pi /p) \theta_1 (2[r-s]\pi /p)} \nonumber \\
&&+\frac{\theta_2 (0)}{\theta_3 (0)} 
\frac{\theta_2 (2s\pi /p)}{\theta_1 (2s\pi /p)}
\left(\frac{\theta_3 (2r\pi /p)}{\theta_1 (2r\pi /p)}
-\frac{\theta_3 (2[r-s]\pi /p)}{\theta_1 (2[r-s]\pi /p)}\right)
\bigg] 
\sum_{j=1}^{p} 
\frac{\theta_2 (z_j)}{\theta_4 (z_j)}~.
\eeqa
\beqa\label{109}
&&\sum_{j=1}^{p} \frac{\theta_1 (z_j)}{\theta_4 (z_{j})}
\left[\frac{\theta_1 (z_{j+r}) \theta_2 (z_{j+s})}
{\theta_4 (z_{j+r}) \theta_4 (z_{j+s})} +
\frac{\theta_1 (z_{j-r}) \theta_2 (z_{j-s})}
{\theta_4 (z_{j-r}) \theta_4 (z_{j-s})}\right]  
= 2
\bigg[\frac{\theta_4 (2r\pi /p) \theta_4 (2[r-s]\pi /p)}
{\theta_1 (2r\pi /p) \theta_1 (2[r-s]\pi /p)} \nonumber \\
&&+\frac{\theta_4 (0)}{\theta_3 (0)}\frac{\theta_4 (2s\pi /p)}
{\theta_1 (2s\pi /p)}
\left(\frac{\theta_3 (2r\pi /p)}{\theta_1 (2r\pi /p)}
-\frac{\theta_3 (2[r-s]\pi /p)}{\theta_1 (2[r-s]\pi /p)}\right)
\bigg] 
\sum_{j=1}^{p} 
\frac{\theta_2 (z_j)}{\theta_4 (z_j)}~.
\eeqa
\beqa\label{110}
&&\sum_{j=1}^{p} \frac{\theta_2 (z_j)}{\theta_4 (z_{j})}
\left[\frac{\theta_1 (z_{j+r}) \theta_1 (z_{j+s})}
{\theta_4 (z_{j+r}) \theta_4 (z_{j+s})} +
\frac{\theta_1 (z_{j-r}) \theta_1 (z_{j-s})}
{\theta_4 (z_{j-r}) \theta_4 (z_{j-s})}\right]  
= 2
\bigg[\frac{\theta_4 (2r\pi /p) \theta_4 (2s\pi /p)}
{\theta_1 (2r\pi /p) \theta_1 (2s\pi /p)} \nonumber \\
&&+\frac{\theta_4 (2[r-s]\pi /p) \theta_4 (0)}
{\theta_1 (2[r-s]\pi /p) \theta_3 (0)}
\left(\frac{\theta_3 (2r\pi /p)}{\theta_1 (2r\pi /p)}
-\frac{\theta_3 (2s\pi /p)}{\theta_1 (2s\pi /p)}\right)
\bigg] 
\sum_{j=1}^{p} 
\frac{\theta_2 (z_j)}{\theta_4 (z_j)}~.
\eeqa
\beqa\label{111}
&&\sum_{j=1}^{p} \frac{\theta_2 (z_j)}{\theta_4 (z_{j})}
\left[\frac{\theta_3 (z_{j+r}) \theta_3 (z_{j+s})}
{\theta_4 (z_{j+r}) \theta_4 (z_{j+s})} +
\frac{\theta_3 (z_{j-r}) \theta_3 (z_{j-s})}
{\theta_4 (z_{j-r}) \theta_4 (z_{j-s})}\right]  
= -2
\bigg[\frac{\theta_2 (2r\pi /p) \theta_2 (2s\pi /p)}
{\theta_1 (2r\pi /p) \theta_1 (2s\pi /p)} \nonumber \\
&&+\frac{\theta_2 (2[r-s]\pi /p) \theta_2 (0)}
{\theta_1 (2[r-s]\pi /p) \theta_3 (0)}
\left(\frac{\theta_3 (2r\pi /p)}{\theta_1 (2r\pi /p)}
-\frac{\theta_3 (2s\pi /p)}{\theta_1 (2s\pi /p)}\right)
\bigg] 
\sum_{j=1}^{p} 
\frac{\theta_2 (z_j)}{\theta_4 (z_j)}~.
\eeqa
\beqa\label{112}
&&\sum_{j=1}^{p} \frac{\theta_2^2 (z_j)}{\theta_4^2 (z_{j})}
\left[\frac{\theta_1 (z_{j+r}) \theta_3 (z_{j+r})}
{\theta_4^2 (z_{j+r})} +
\frac{\theta_1 (z_{j-r}) \theta_3 (z_{j+r})}
{\theta_4^2 (z_{j-r})}\right] \nonumber \\ 
&&= -2 \frac{\theta_2^3 (0) \theta_2 (2r\pi /p)}
{\theta_3^2 (0) \theta_1^2 (2r\pi /p)}
\left[\frac{\theta_4 (2r\pi /p)}{\theta_4 (0)}
+\frac{\theta_2 (0) \theta_3^2 (2r\pi /p)}
{\theta_3^2 (0) \theta_2 (2r\pi /p)}\right]
\sum_{j=1}^{p} \frac{\theta_1 (z_j) \theta_3 (z_j)}{\theta_4^2 (z_j)}~.
\eeqa
\beqa\label{113}
&&\sum_{j=1}^{p} \frac{\theta_1 (z_j) \theta_2 (z_j)}{\theta_4^2 (z_{j})}
\left[\frac{\theta_2 (z_{j+r}) \theta_3 (z_{j+r})}
{\theta_4^2 (z_{j+r})} +
\frac{\theta_2 (z_{j-r}) \theta_3 (z_{j-r})}
{\theta_4^2 (z_{j-r})}\right] \nonumber \\ 
&&= -2 \frac{\theta_2^2 (0) \theta_3 (2r\pi /p)}
{\theta_3 (0) \theta_1^2 (2r\pi /p)}
\left[\frac{\theta_4 (2r\pi /p)}{\theta_4 (0)} 
+\frac{\theta_2 (2r\pi /p)}
{\theta_2 (0)}\right]
\sum_{j=1}^{p} \frac{\theta_1 (z_j) \theta_3 (z_j)}{\theta_4^2 (z_j)}~.
\eeqa
\beq\label{114}
\sum_{j=1}^{p} \frac{\theta_2^2 (z_j) \theta_3 (z_j)}{\theta_4^3 (z_{j})}
\left[\frac{\theta_1 (z_{j+r})}
{\theta_4 (z_{j+r})} +
\frac{\theta_1 (z_{j-r})}
{\theta_4 (z_{j-r})}\right] 
= 2 \frac{\theta_2^3 (0) \theta_2 (2r\pi /p) \theta_3 (2r\pi /p)}
{\theta_3^3 (0) \theta_1^2 (2r\pi /p)}
\sum_{j=1}^{p} \frac{\theta_1 (z_j) \theta_3 (z_j)}{\theta_4^2 (z_j)}~.
\eeq
\beqa\label{115}
&&\sum_{j=1}^{p} \frac{\theta_1 (z_j) \theta_2 (z_j) \theta_3 (z_j)}
{\theta_4^3 (z_{j})}
\left[\frac{\theta_2^3 (z_{j+r})}
{\theta_4^3 (z_{j+r})} +
\frac{\theta_2^3 (z_{j-r})}
{\theta_4^3 (z_{j-r})}\right] \nonumber \\ 
&&=  \frac{2\theta_2^2 (0) \theta_2^2 (2r\pi /p)}
{\theta_1^4(2r\pi /p)}
\bigg[\frac{\theta_4^2 (2r\pi /p)}{\theta_4^2 (0)}
+\frac{\theta_2^2 (0) \theta_3^2 (2r\pi /p) \theta_4^2 (2r\pi /p)}
{\theta_3^2 (0) \theta_4^2 (0) \theta_2^2 (2r\pi /p)}+
\frac{\theta_3^2 (2r\pi /p)}
{\theta_3^2 (0)} \nonumber \\
&&+3\frac{\theta_2 (0) \theta_4 (2r\pi /p) \theta_3^2 (2r\pi /p)}
{\theta_3^2 (0) \theta_2 (2r\pi /p) \theta_4 (0)}\bigg] 
\sum_{j=1}^{p} \frac{\theta_1 (z_j) \theta_3 (z_j)}{\theta_4^2 (z_j)}~.
\eeqa
\beqa\label{116}
&&\sum_{j=1}^{p} \frac{\theta_1 (z_j) \theta_3 (z_j)}{\theta_4^2 (z_{j})}
\left[\frac{\theta_2 (z_{j+r}) \theta_2 (z_{j+s})}
{\theta_4 (z_{j+r}) \theta_4 (z_{j+s})} +
\frac{\theta_2 (z_{j-r}) \theta_2 (z_{j-s})}
{\theta_4 (z_{j-r}) \theta_4 (z_{j-s})}\right] \nonumber \\
&&= -2 \frac{\theta_2^4 (0) \theta_3 (2r\pi /p) \theta_3 (2s\pi /p)}
{\theta_3^4 (0) \theta_1 (2r\pi /p) \theta_1 (2s\pi /p)}
\sum_{j=1}^{p} \frac{\theta_1 (z_j) \theta_3 (z_j)}{\theta_4^2 (z_j)}~.
\eeqa
\beqa\label{117}
&&\sum_{j=1}^{p} \frac{\theta_1 (z_j) \theta_3 (z_j)}{\theta_4^2 (z_{j})}
\left[\frac{\theta_3 (z_{j+r}) \theta_3 (z_{j+s})}
{\theta_4 (z_{j+r}) \theta_4 (z_{j+s})} +
\frac{\theta_3 (z_{j-r}) \theta_3 (z_{j-s})}
{\theta_4 (z_{j-r}) \theta_4 (z_{j-s})}\right] \nonumber \\
&&= -2 \frac{\theta_2^4 (0)}{\theta_3^4 (0)} 
\frac{\theta_2 (2r\pi /p) \theta_2 (2s\pi /p)}
{\theta_1 (2r\pi /p) \theta_1 (2s\pi /p)}
\sum_{j=1}^{p} \frac{\theta_1 (z_j) \theta_3 (z_j)}{\theta_4^2 (z_j)}~.
\eeqa
\beqa\label{118}
&&\sum_{j=1}^{p} \frac{\theta_1 (z_j) \theta_3 (z_j)}{\theta_4^2 (z_{j})}
\left[\frac{\theta_1 (z_{j+r}) \theta_1 (z_{j+s})}
{\theta_4 (z_{j+r}) \theta_4 (z_{j+s})} +
\frac{\theta_1 (z_{j-r}) \theta_1 (z_{j-s})}
{\theta_4 (z_{j-r}) \theta_4 (z_{j-s})}\right] \nonumber \\
&&= 2 \frac{\theta_2^4 (0) \theta_4 (2r\pi /p) \theta_4 (2s\pi /p)}
{\theta_3^4 (0) \theta_1 (2r\pi /p) \theta_1 (2s\pi /p)}
\sum_{j=1}^{p} \frac{\theta_1 (z_j) \theta_3 (z_j)}{\theta_4^2 (z_j)}~.
\eeqa
\beqa\label{119}
&&\sum_{j=1}^{p} \frac{\theta_2 (z_j) \theta_3 (z_j)}{\theta_4^2 (z_{j})}
\left[\frac{\theta_1 (z_{j+r}) \theta_2 (z_{j+s})}
{\theta_4 (z_{j+r}) \theta_4 (z_{j+s})} +
\frac{\theta_1 (z_{j-r}) \theta_2 (z_{j-s})}
{\theta_4 (z_{j-r}) \theta_4 (z_{j-s})}\right] \nonumber \\
&&= -2 \frac{\theta_2^4 (0) \theta_4 (0) \theta_4 (2r\pi /p) 
\theta_3 (2s\pi /p)}
{\theta_3^5 (0) \theta_1 (2r\pi /p) \theta_1 (2s\pi /p)}
\sum_{j=1}^{p} \frac{\theta_1 (z_j) \theta_3 (z_j)}{\theta_4^2 (z_j)}~.
\eeqa
\beqa\label{120}
&&\sum_{j=1}^{p} \frac{\theta_1^2 (z_j) \theta_3^2 (z_j)}
{\theta_4^4 (z_{j})}\left[\frac{\theta_2 (z_{j+r})}
{\theta_4 (z_{j+r})} +
\frac{\theta_2 (z_{j-r})}
{\theta_4 (z_{j-r})}\right]  
= -2 \frac{\theta_2 (2r\pi /p) \theta_4 (2r\pi /p)}
{\theta_1^2 (2r\pi /p)}
\sum_{j=1}^{p} \frac{\theta_2^3 (z_j)}{\theta_4^3 (z_j)} \nonumber \\
&&+2\frac{\theta_2^3 (0) \theta_2 (2r\pi /p) \theta_4 (2r\pi /p)}
{\theta_4 (0) \theta_3^2 (0) \theta_1^2 (2r\pi /p)}
\bigg[1+
\frac{\theta_3^2 (0) \theta_4^2 (0) \theta_2^2 (2r\pi /p)}
{\theta_1^2 (2r\pi /p) \theta_2^4 (0)} \nonumber \\
&&-\frac{\theta_3^2 (0) \theta_4 (0) \theta_2 (2r\pi /p) \theta_4 (2r\pi /p)}
{\theta_1^2 (2r\pi /p) \theta_2^3 (0)}\bigg]
\sum_{j=1}^{p} \frac{\theta_2 (z_j)}{\theta_4 (z_j)}~.
\eeqa
\beqa\label{121}
&&\sum_{j=1}^{p} \frac{\theta_2^3 (z_j)}
{\theta_4^3 (z_{j})}\left[\frac{\theta_2^2 (z_{j+r})}
{\theta_4^2 (z_{j+r})} +
\frac{\theta_2^2 (z_{j-r})}
{\theta_4^2 (z_{j-r})}\right]  
= -2 \frac{\theta_4^2 (0) \theta_3^2 (2r\pi /p)}
{\theta_3^2 (0)\theta_1^2 (2r\pi /p)}
\sum_{j=1}^{p} \frac{\theta_2^3 (z_j)}{\theta_4^3 (z_j)} \nonumber \\
&&+2\frac{\theta_3^4 (0) \theta_4^2 (2r\pi /p)}
{\theta_4^2 (0) \theta_1^4 (2r\pi /p)}
\bigg[\frac{\theta_2^2 (2r\pi /p)}{\theta_2^2 (0)}+
\frac{\theta_3^2 (2r\pi /p)}
{\theta_3^2 (0)} \nonumber \\
&&+\frac{\theta_4^2 (0) \theta_2^2 (2r\pi /p) \theta_3^2 (2r\pi /p)}
{\theta_4^2 (2r\pi /p) \theta_2^2 (0) \theta_3^2 (0)}
-3\frac{\theta_2 (2r\pi /p) \theta_4 (0) \theta_3^2 (2r\pi /p)}
{\theta_3^2 (0) \theta_2 (0) \theta_4 (2r\pi /p)}\bigg]
\sum_{j=1}^{p} \frac{\theta_2 (z_j)}{\theta_4 (z_j)}~.
\eeqa

\section{Comments and discussion.}

We shall now give some general comments about extending the above results in
several directions.

{\bf (i) Identities for auxiliary functions:} Until now we have discussed 
identities for the three basic ratios 
$\theta_1(z) /\theta_4 (z)$, $\theta_2(z) /\theta_4 (z)$,
$\theta_3(z) /\theta_4 (z)$. What about the identities for the remaining nine
ratios like say $\theta_1 (z) /\theta_2 (z)$? These are readily obtained by
making use of the identities coming from increasing $z$ by the half
periods $\pi /2$, $\pi \tau /2$ and $\pi (1+\tau) /2$. For example, 
using \cite{law}
\beq\label{122}
\frac{\theta_1 (z)}{\theta_2 (z)} = 
-i\frac{\theta_3 (z+\frac{\pi}{2}[1+\tau])}
{\theta_4 (z+\frac{\pi}{2}[1+\tau])}~,
\eeq
we immediately obtain identities for the ratio $\theta_1 (z) /\theta_2 (z)$.

{\bf (ii) Identities for shifts in units of $\pi \tau /p$ or 
$\pi (1-\tau) /p$:} So far we have
focused our attention on identities 
involving ratios of Jacobi theta functions evaluated at points separated by
gaps of $T /p$ with $T$ real (and equal to $\pi$ or $2\pi$). But since the 
ratios of theta functions are also periodic with period $T \tau /p$ as well as
$T(1-\tau) /p$, we can convert each of our identity to another one involving
points separated either by gaps of $T \tau /p$ or $T(1-\tau) /p$. In fact
this is easily done using the modular transformations \cite{law}
\beq\label{123}
\frac{\theta_1 (z,\tau_1)}{\theta_4 (z,\tau_1)} =
-i\frac{\theta_1 (\tau z,\tau)}{\theta_2 (\tau z,\tau)}~, 
~\frac{\theta_2 (z,\tau_1)}{\theta_4 (z,\tau_1)} =
\frac{\theta_4 (\tau z,\tau)}{\theta_2 (\tau z,\tau)}~, 
~\frac{\theta_3 (z,\tau_1)}{\theta_4 (z,\tau_1)} =
-i\frac{\theta_3 (\tau z,\tau)}{\theta_2 (\tau z,\tau)}~, 
\eeq
\beq\label{124}
\frac{\theta_1 (z,\tau_3)}{\theta_4 (z,\tau_3)} =
\frac{\theta_1 ([1-\tau] z,\tau)}{\theta_4 ([1-\tau] z,\tau)}~, 
~\frac{\theta_2 (z,\tau_3)}{\theta_4 (z,\tau_3)} =
\frac{\theta_3 ([1-\tau] z,\tau)}{(i)^{1/2} \theta_4 ([1-\tau] z,\tau)}~, 
~\frac{\theta_3 (z,\tau_3)}{\theta_4 (z,\tau_3)} =
\frac{\theta_2 ([1-\tau] z,\tau)}{(i)^{1/2} \theta_4 ([1-\tau] z,\tau)}~, 
\eeq
where $\tau_1 = -1/\tau$ while $\tau_3 = \tau /(1-\tau)$ which correspond to
changing the modular parameter $m$ to $1-m$ and $1/m$ respectively.

{\bf (iii) Identities for products of ratios of Jacobi theta functions:} 
In previous discussions, we have evaluated Jacobi theta functions at
points separated by gaps of $T/p, T\tau /p$ or $T(1-\tau) /p$. An obvious 
question is if we can also obtain identities for products of ratios of theta
functions like say $\theta_1 \theta_2 /\theta_3 \theta_4$. The answer is yes
and in fact these can be easily obtained from the identities derived in this
paper by noting the relations
\beq\label{125}
\frac{\theta_2 (z) \theta_3 (z)}{\theta_1 (z) \theta_4 (z)}
=i\frac{\theta_3 (0) \theta_3 (2z+\pi \tau /2)
+ \theta_2 (0)\theta_2 (2z+ \pi \tau /2)} 
{\theta_4 (0) \theta_4 (2z+\pi \tau /2)}~.
\eeq
\beq\label{126}
\frac{\theta_1 (z) \theta_3 (z)}{\theta_2 (z) \theta_4 (z)}
=\frac{\theta_2 (0) \theta_1 (2z+\pi \tau /2)
-i \theta_4 (0)\theta_3 (2z+ \pi \tau /2)} 
{\theta_3 (0) \theta_4 (2z+\pi \tau /2)}~.
\eeq
\beq\label{127}
\frac{\theta_1 (z) \theta_2 (z)}{\theta_3 (z) \theta_4 (z)}
=\frac{\theta_3 (0) \theta_1 (2z+\pi \tau /2)
-i \theta_4 (0)\theta_2 (2z+ \pi \tau /2)} 
{\theta_2 (0) \theta_4 (2z+\pi \tau /2)}~.
\eeq
\beq\label{128}
\frac{\theta_2 (z) \theta_4 (z)}{\theta_1 (z) \theta_3 (z)}
=\frac{\theta_2 (0) \theta_1 (2z+\pi \tau /2)
+i \theta_4 (0)\theta_3 (2z+ \pi \tau /2)} 
{\theta_3 (0) \theta_4 (2z+\pi \tau /2)}~.
\eeq
\beq\label{129}
\frac{\theta_1 (z) \theta_4 (z)}{\theta_2 (z) \theta_3 (z)}
=i\frac{\theta_2 (0) \theta_2 (2z+\pi \tau /2)
- \theta_3 (0)\theta_3 (2z+ \pi \tau /2)} 
{\theta_3 (0) \theta_4 (2z+\pi \tau /2)}~.
\eeq
\beq\label{130}
\frac{\theta_3 (z) \theta_4 (z)}{\theta_1 (z) \theta_2 (z)}
=\frac{\theta_3 (0) \theta_1 (2z+\pi \tau /2)
+i \theta_4 (0)\theta_2 (2z+ \pi \tau /2)} 
{\theta_2 (0) \theta_4 (2z+\pi \tau /2)}~.
\eeq

Another interesting question to which we do not know the answer is, just as the ratios of Jacobi theta 
functions (which usually occur with genus one in string theory) satisfy various
identities, do the higher genus theta functions also satisfy similar 
identities?  

One of us (US) gratefully acknowledges partial support of this research from 
the U.S. Department of Energy.


\newpage

\begin{thebibliography}{99}

\bibitem{ksjmp}A. Khare, U. Sukhatme, {\it Cyclic Identities Involving Jacobi 
Elliptic Functions}, J. Math. Phys. {\bf 43}, 3798 (2002). 
%
\bibitem{klsjmp}A. Khare, A. Lakshminarayan and U. Sukhatme, 
{\it Cyclic Identities for Jacobi Elliptic and Related 
Functions}, J. Math. Phys. {\bf 44}, 1822 (2002). 
%
\bibitem{klsmp}A. Khare, A. Lakshminarayan and U. Sukhatme, 
{\it Cyclic Identities Involving Jacobi Elliptic Functions. II},
arXiv:math-ph/0207019. 
%
\bibitem{abr} For the properties of Jacobi elliptic functions, see, for example, M. Abramowitz and I. Stegun, {\it Handbook of Mathematical Functions} (Dover, 1964);
I. S. Gradshteyn and I. M. Ryzhik, {\it Table of Integrals, Series
and Products} (Academic Press, 1980); P. F. Byrd and M. D.
Friedman, {\it Handbook of Elliptic Integrals for Engineers and
Physicists} (Springer Verlag, 1954).
%
\bibitem{ksprl} A. Khare and U. Sukhatme, {\it Linear Superposition in 
Nonlinear Equations}, Phys. Rev. Lett. {\bf 88}, 244101 (2002).
%
\bibitem{cks} F. Cooper, A. Khare and U. Sukhatme, 
{\it Periodic Solutions of Nonlinear Equations
Obtained by Linear Superposition}, J. Phys. A: Math. Gen. {\bf 35}, 
10085 (2001).
%
\bibitem{law}  D. F. Lawden, {\it Elliptic Functions and Applications} (Springer, 1989).
%
\bibitem{klspr}A. Khare, A. Lakshminarayan and U. Sukhatme, 
{\it Local Identities Involving Jacobi Elliptic Functions},
arXiv:math-ph/0306028. 
%


\end{thebibliography}
\end{document}